\documentclass[aps,prb,superscriptaddress,twocolumn]{revtex4-1}
\usepackage{amsmath}
\usepackage{amssymb}
\usepackage{amsthm}
\usepackage{graphicx}
\usepackage{graphics}
\usepackage{subfigure}
\usepackage{bm}
\usepackage{hyperref}
\usepackage{rotating}

\newcommand{\<}{\langle}
\renewcommand{\>}{\rangle}

\def\beas{\begin{eqnarray*}}
\def\eeas{\end{eqnarray*}}
\def\bea{\begin{eqnarray}}
\def\eea{\end{eqnarray}}
\def\be{\begin{equation}}
\def\ee{\end{equation}}

\newcommand{\down}{\downarrow}

\newcommand{\bpm}{\begin{pmatrix}}
\newcommand{\epm}{\end{pmatrix}}
\newcommand{\bmm}{\begin{matrix}}
\newcommand{\emm}{\end{matrix}}
\newcommand{\up}{\uparrow}

\begin{document}

\title{Classification and analysis of two dimensional abelian fractional topological insulators}

\author{Michael Levin}
\affiliation{Condensed Matter Theory Center, Department of Physics, University of Maryland, College Park, Maryland 20742, USA}
\author{Ady Stern}
\affiliation{Department of Condensed Matter Physics, Weizmann
Institute of Science, Rehovot 76100, Israel}

%\date{\today}

\begin{abstract}
We present a general framework for analyzing fractionalized, time reversal invariant
electronic insulators in two dimensions. The framework applies to all insulators whose
quasiparticles have abelian braiding statistics. First, we construct the most general
Chern-Simons theories that can describe these states. We then derive a criterion for when
these systems have protected gapless edge modes -- that is, edge modes that cannot be gapped out without
breaking time reversal or charge conservation symmetry. The systems with protected edge modes can be regarded as
fractionalized analogues of topological insulators. We show that previous examples of 2D fractional topological
insulators are special cases of this general construction. As part of our derivation, we define the concept
of ``local Kramers degeneracy'' and prove a local version of Kramers theorem.
\end{abstract}

%\pacs{}

\maketitle

% ----------------------------------------------------------------
\section{Introduction}
Recently it was realized that there are two kinds of
two dimensional (2D) time reversal invariant band insulators: topological insulators and trivial
insulators. These two types of insulators are distinguished by the fact that topological
insulators have robust gapless edge modes while trivial insulators do
not.\cite{KaneMele,KaneMele2,BernevigZhang,HasanKaneRMP}

While much of the work on topological insulators has focused on non-interacting
or weakly interacting systems, it is natural to wonder whether similar physics
can occur in systems with strong interactions. Such strongly interacting insulators can
be divided into two classes: systems that can be adiabatically deformed into
(non-interacting) band insulators without closing the bulk gap, and systems that
cannot. At some level the first case is understood: it is known\cite{FuKanePump} that the gapless
edge modes of a topological insulator are stable to arbitrary local interactions as
long as time reversal symmetry and charge conservation are not broken (explicitly
or spontaneously). Here we will consider the second case: strongly interacting, time reversal
invariant electron systems whose ground state cannot be adiabatically connected
to a band insulator. These systems are typically fractionalized in the sense
that they have quasiparticle excitations with fractional charge and fractional statistics.

Previous work has focused on particular classes of these fractionalized insulators. Ref. [\onlinecite{LevinStern}] 
considered toy models\cite{BernevigZhang} where spin-up and spin-down electrons each
form decoupled abelian fractional quantum Hall states with opposite chiralities.\cite{Freedmanetal}
The authors showed that some of these systems have protected edge modes, while
some do not. More precisely, the authors found that these models have a protected edge mode if
and only if $\sigma_{sH}/e^*$ is odd, where $e^*$ is the elementary charge (in units of $e$)
and $\sigma_{sH}$ is the spin-Hall conductivity (in units of $e/2\pi$). The two kinds of
insulators were dubbed ``fractional topological insulators'' and ``fractional
trivial insulators'' by analogy to non-interacting topological
and trivial insulators. Along the same lines, Refs. [\onlinecite{Neupertetal}] and [\onlinecite{Santosetal}] 
considered generalizations of the above toy models where spin-up and spin-down electrons form \emph{correlated} 
fractional quantum Hall states. Finally, Ref. [\onlinecite{LevinBurnellKoch}] considered a somewhat different class of time 
reversal invariant fractionalized insulators which were realized by exactly soluble lattice models. 
In all these cases, it was found that some of the fractionalized insulators have protected edge modes while some do not.

Taken together, these papers have analyzed a large class of time reversal invariant fractionalized
insulators. However, they have not exhausted all the possibilities. Here we give a more
complete analysis of these systems. We consider \emph{general} time reversal invariant 2D electronic
insulators with abelian quasiparticle statistics. We then derive
a criterion for when these systems have protected gapless edge modes -- that is, edge modes
that cannot be gapped out without breaking time reversal or charge conservation symmetry, explicitly
or spontaneously. We call the states with protected edge modes ``fractional topological insulators'', 
following the terminology of Ref. [\onlinecite{LevinStern}]. 

Our analysis proceeds in two steps. First, we find the most general Chern-Simons
theory that can describe a time reversal invariant, abelian electronic insulator. Then, we investigate the stability
of the edge modes of these states, using both a microscopic approach and a macroscopic flux insertion argument.
Much of the edge stability analysis can be regarded as a direct generalization of the arguments of Ref.
[\onlinecite{LevinStern}]. However, there are a number of additional features coming from
the more detailed exposition of this paper. In particular, we give a precise definition of
``local Kramers degeneracy'' and a proof of a local analogue of Kramers theorem.

This paper is organized as follows. In section \ref{CSdescsec}, we find a Chern-Simons description
of general time reversal invariant insulators with abelian quasiparticle statistics.
In section \ref{microedgesec}, we analyze the stability of the edges of these insulators
using a microscopic approach. In section \ref{fluxargsec}, we analyze the edge stability 
using a flux insertion argument and in the process give a proof of a local analogue of Kramers theorem. 
The appendix contains some of the more technical calculations.

\section{Chern-Simons theories for time reversal invariant insulators} \label{CSdescsec}
\subsection{The time-reversal breaking case}
First, we review the Chern-Simons description
for abelian insulators without time reversal symmetry.\cite{WenBook,WenReview,WenKmatrix} In the next section, we discuss
how this formalism needs to be modified to describe the time reversal symmetric case.
To begin, recall that a general abelian insulator can be described by a $p$-component $U(1)$ Chern-Simons theory of the form
\begin{equation}
L_{B} = \frac{\mathcal{K}_{IJ}}{4\pi} \epsilon^{\lambda \mu \nu} a_{I \lambda} \partial_\mu a_{J \nu}
- \frac{1}{2\pi} \tau_I \epsilon^{\lambda \mu \nu} A_{\lambda} \partial_\mu a_{I \nu}
\label{kmat}
\end{equation}
where $\mathcal{K}$ is a $p \times p$ symmetric, non-degenerate integer matrix and $\tau$ is a $p$ component
integer vector. We will refer to $\mathcal{K}$ as the ``$K$-matrix'' and $\tau$ as the ``charge vector.''
The first term in $L_B$ describes the different degrees of freedom in the insulator while the second term
describes their coupling to the electromagnetic vector potential $A_\lambda$. In general, $L_B$ can also
contain additional terms such as $(\partial_\mu a_{I\nu} - \partial_\nu a_{I\mu})^2$,
but these terms do not affect the basic topological features of the insulator -- our main interest here.

In this formalism, the quasiparticle excitations in the insulator are described
by coupling $L_B$ to bosonic particles which carry \emph{integer} gauge charge $l_I$
under each of the gauge fields $a_I$. Thus, each quasiparticle excitation corresponds to a
$p$ component integer vector $l$. The physical electric charge of each excitation is
given by
\begin{equation}
q_l = \frac{1}{2\pi} l^T \mathcal{K}^{-1} \tau
\label{charge}
\end{equation}
while the mutual statistics associated with braiding one particle around another is given by
\begin{equation}
\theta_{ll'} = 2\pi l^T \mathcal{K}^{-1} l'
\label{stat}
\end{equation}
The statistical phase associated with exchanging two particles is $\theta_l = \theta_{ll}/2$.
The quasiparticle excitations which are ``local''  -- that is, composed out of the
constituent electrons -- correspond to vectors $l$ of the form $l = \mathcal{K} \Lambda$
where $\Lambda$ is an integer $p$ component vector.

If we don't require time reversal symmetry, the $K$-matrix and charge vector are unconstrained except for two requirements. First, 
we must have
\begin{equation}
\tau_{I} \equiv \mathcal{K}_{II} \text{ (mod $2$)}
\label{paritycond}
\end{equation}
To derive this constraint, consider the statistics and charges of the local excitations
with $l = K \Lambda$. From (\ref{charge}), (\ref{stat}) we can see that the excitation
corresponding to $l = \mathcal{K} \Lambda$ carries electric charge $\Lambda^T \tau$ and has exchange statistics 
$\theta = \pi \Lambda^T \mathcal{K} \Lambda$. At the same time, we know that local excitations 
with even charge must be bosons, while those with odd charge must be fermions, since they are 
composed out of electrons. Combining these two observations yields (\ref{paritycond}).

The second constraint on $\mathcal{K},\tau$ is that 
\begin{equation}
\text{gcd}(\tau_1,...,\tau_p) = 1
\label{taucond}
\end{equation}
where the notation ``gcd'' denotes the greatest common divisor of these integers.
This constraint follows from the requirement that at least one local excitation has the charge
of an electron. (Readers familiar with the ``strong pairing'' fractional quantum Hall state 
may worry that the condition (\ref{taucond}) is too strict, since this state is traditionally described 
using a Chern-Simons theory with $\mathcal{K} = 8, \tau = 2$, apparently violating (\ref{taucond}). 
However, this state can be described equally well by the $3 \times 3$ $K$-matrix with diagonal 
elements $8,1,-1$, and the charge vector $\tau = (2, 1, 1)$, thus satisfying (\ref{taucond}). The same 
trick can be used to find a Chern-Simons representation satisfying (\ref{taucond}) for any abelian 
insulator built out of electrons).

\subsection{Including time reversal symmetry}
We now show how to extend the above formalism to describe abelian insulators with the
additional symmetry of time reversal. The first step is to describe the action of time
reversal on the field theory (\ref{kmat}). We will assume that the time reversal transformation
$\mathcal{T}$ is of the form
\begin{equation*}
a_I \rightarrow T_{IJ} a_J
\end{equation*}
where $T$ is  an integer $p \times p$ matrix. More precisely, because time reversal
acts differently on spatial and temporal coordinates, we will assume that
\begin{equation}
a_{I \mu} \rightarrow \pm T_{IJ} a_{J \mu}
\label{gent}
\end{equation}
where the sign is $+$ for $\mu = 1,2$ and $-$ for $\mu = 0$. To see where this transformation law comes from, recall
that, in our formalism, the quasiparticle excitations are described by sources carrying integer gauge charge. 
Therefore, integer gauge charge must transform into integer gauge charge under time reversal.
Equation \ref{gent} follows immediately.

In order for the action (\ref{kmat}) to be invariant under the
time reversal transformation (\ref{gent}), we must have
\begin{eqnarray}
T^T \mathcal{K} T &=& -\mathcal{K} \label{tkcond}\\
T \tau &=& \tau \label{tchgcond}
\end{eqnarray}
Similarly, the requirement that $\mathcal{T}^2 = (-1)^{N_e}$, where $N_e$ is the total number of electrons in the
system, implies that
\begin{equation}
T^2 = \bf{1}_P  \label{t2cond}
\end{equation}
where $\bf{1}_P$ is the $p \times p$ identity matrix.

We now come to an important point: Eq. \ref{gent} does not completely specify the action of time reversal. The
problem is that (\ref{gent}) only tells us how time reversal acts on operators like
$\epsilon^{\lambda \mu \nu} \partial_\mu a_{I \nu}$. These operators are all electrically neutral -- that is,
they preserve the total electric charge in the system. To complete our description of time reversal
we also have to specify the transformation properties of \emph{charged} operators, like electron creation
and annihilation operators. 

Unfortunately, the Chern-Simons theory is not a convenient framework for describing these operators. The 
reason is that these charged operators correspond to magnetic monopole instantons in the 
gauge fields $a_I$. Thus, if we want to complete our description within the Chern-Simons theory, 
we need to introduce additional notation and formalism to describe these instantons and their
transformation properties under $\mathcal{T}$. 

We can avoid these (technical) complications by instead describing the action of time reversal at the \emph{edge}.
According to the usual bulk-edge correspondence for abelian Chern-Simons theories, a valid edge theory for
(\ref{kmat}) is given by\cite{WenReview,WenBook}
\begin{eqnarray}
L &=& \frac{1}{4\pi} (\mathcal{K}_{IJ} \partial_x \Phi_I \partial_t \Phi_J
- \mathcal{V}_{IJ} \partial_x \Phi_I \partial_x \Phi_J) \nonumber \\
&+& \frac{1}{2\pi} \epsilon^{\mu \nu} \tau_I \partial_\mu \Phi_I A_\nu
\label{kedge}
\end{eqnarray}
where $\mathcal{V}$ is a $p \times p$ velocity matrix. Here, we are using a
normalization convention where the quasiparticle excitations are of the form
$e^{i l^T \Phi}$ where $l$ is an integer $p$ component vector. Local operators (i.e. operators composed out of products of
electron creation and annihilation operators) are of the form $e^{i\Theta(\Lambda)}$ with
$\Theta(\Lambda)\equiv \Lambda^T \mathcal{K}\Phi$ and
$\Lambda$ an integer vector.

Translating the transformation law (\ref{gent}) into the edge theory language gives
\begin{equation}
\mathcal{T}^{-1} (\partial_x \Phi_I) \mathcal{T} = T_{IJ} \partial_x \Phi_J
\label{tdx}
\end{equation}
along with a similar expression for the temporal derivative $\partial_t \Phi_I$.
The missing component in our description is now clear: equation \ref{tdx} 
does not determine the transformation properties for the local operators
$e^{i\Theta(\Lambda)}$, i.e. the creation and annihilation operators for
electrons.

We can see how to complete our description if we rewrite (\ref{tdx}) as
\begin{equation*}
\mathcal{T}^{-1} \Phi_I \mathcal{T} = T_{IJ} \Phi_J + \text{const.}
\end{equation*}
or equivalently
\begin{equation}
\mathcal{T}^{-1} \Phi_I \mathcal{T} = T_{IJ} \Phi_J + \pi \mathcal{K}^{-1}_{IJ} \chi_J
\label{gentedge}
\end{equation}
where $\chi$ is some $p$-component real vector. Clearly, in order to complete our description of
time reversal we need to specify $\chi$. We will refer to $\chi$ as the ``time reversal vector''.

Like $T$, the vector $\chi$ cannot be chosen arbitrarily: it is constrained by the requirement
that electron creation and annihilation operators are odd under $\mathcal{T}^2$. This requirement implies that
\begin{equation}
\mathcal{T}^{-2} e^{i\Theta(\Lambda)} \mathcal{T}^2 = e^{i\Theta(\Lambda)} \cdot (-1)^{\Lambda^T \tau}
\label{tsquared}
\end{equation}
which is equivalent to the constraint that
\begin{equation}
(1-T^T)\chi \equiv \tau \text{ (mod $2$)}
\label{tchicond}
\end{equation}
We note that Klein factors are not relevant to this discussion of time reversal properties and therefore we have
dropped them for clarity. In fact, we will see that Klein factors do not affect our later analysis 
either and can be safely ignored throughout this paper.

In summary, we have shown that a general abelian time reversal invariant insulator is described by the data $(\mathcal{K},\tau,T,\chi)$.
The $K$-matrix $\mathcal{K}$ and charge vector $\tau$ specify the quasiparticle braiding
statistics and charges, while the time reversal matrix $T$ and vector $\chi$ describe the action of time reversal.
This data must satisfy the conditions (\ref{paritycond}-\ref{taucond}), (\ref{tkcond}-\ref{t2cond}) and 
(\ref{tchicond}) in order to be self-consistent.

\subsection{Explicit parameterization of time reversal invariant abelian Chern-Simons theories}
An important point is that not all $(\mathcal{K},\tau,T,\chi)$ correspond to physically
distinct Chern-Simons theories. For example, $(\mathcal{K},\tau,T,\chi)$ is physically equivalent to
\begin{eqnarray}
\mathcal{K}' &=& (U^{-1})^T \mathcal{K} U^{-1} \nonumber \\
\tau' &=& (U^{-1})^T \tau \nonumber \\
T' &=& U T U^{-1} \nonumber \\
\chi' &=& (U^{-1})^T \chi
\label{uequiv}
\end{eqnarray}
for any integer matrix $U$ with determinant $\pm 1$. This equivalence can be derived by making the 
change of variables $a' = U a$, $\Phi' = U \Phi$ in the bulk/edge theories (\ref{kmat}),(\ref{kedge}).
Another important equivalence is the one between $\chi$ and 
\begin{equation}
\chi' = \chi + \frac{1}{\pi} \mathcal{K} (1 - T) \xi
\label{chiequiv}
\end{equation}
for any real vector $\xi$, which follows from the field redefinition $\Phi' = \Phi + \xi$.
A final example is that $\chi$ is physically equivalent to
\begin{equation}
\chi' = \chi + 2 v
\label{chiequiv2}
\end{equation}
for any integer vector $v$. To see where this equivalence comes from, note that the only place where
$\chi$ enters into the physical description of the edge is in the transformation law for
$e^{i \Theta(\Lambda)}$:
\begin{equation}
\mathcal{T}^{-1} e^{i \Theta(\Lambda)} \mathcal{T} = e^{i \Theta(T \Lambda) - i \pi \Lambda^T \chi}
\end{equation}
We can see that $\chi$ appears in the combination $e^{i\pi \Lambda^T \chi}$,
which only depends on the value of $\chi$ modulo $2$.

We are now in a position to explicitly write down all time reversal invariant abelian 
Chern-Simons theories. Indeed, according to the above analysis, it suffices to find all
$(\mathcal{K}, \tau, T, \chi)$ satisfying conditions (\ref{paritycond}-\ref{taucond}), 
(\ref{tkcond}-\ref{t2cond}).
This problem can be solved by straightforward linear algebra as we demonstrate in appendix \ref{genabelapp}. 
We find that the most general solution, up to the transformations (\ref{uequiv}), (\ref{chiequiv}) and (\ref{chiequiv2}), 
is of the form
\begin{align}
\mathcal{K} = \bpm 0 & A & B & B \\
		   A^T & 0 & C & -C \\
		   B^T & C^T & K & W \\
		   B^T & -C^T & W^T & -K \epm \ , \ \tau = \bpm 0 \\ t' \\ t \\ t \epm
\label{gensolK}
\end{align}
\begin{align}
T = \bpm -\bf{1}_M & 0 & 0 & 0 \\
	     0 & \bf{1}_{M} & 0 & 0 \\
	     0 & 0 & 0 & \bf{1}_{N-M} \\
	     0 & 0 & \bf{1}_{N-M} & 0 \epm  \ , \ \chi = \bpm x \\ 0 \\ 0 \\ t \epm
\label{gensolT}
\end{align}
Here, the matrix $A$ is of dimension $M \times M$, while the matrices $B,C$ are of dimension $M \times (N-M)$. The matrices $K,W$
are both of dimension $(N-M)\times(N-M)$. Similarly, $t'$ is of dimension $M$ and $t$ is of dimension $(N-M)$. Finally, the 
vector $x$ is some $(N-M)$ dimensional vector of $1$'s and $0$'s. We note that that the total dimension of $\mathcal{K}$ is 
$2N \times 2N$ so the above solution corresponds to $p = 2N$ in our previous notation.

There are only a few constraints on $(A,B,C,K,W,t,t',x)$. First, $W$ must be antisymmetric: $W = -W^T$. This requirement
follows from time reversal invariance (\ref{tkcond}). Second, $t'$ must be even-valued. This constraint comes from the 
condition (\ref{paritycond}) that the insulator is composed out of electrons. For the same reason, the parity of $t_I$ 
must match that of $K_{II}$, but can be either even or odd. Finally, the greatest common factor of $\{\tau_I\}$ must be 
$1$ according to (\ref{taucond}).

We would like to mention that while every time reversal invariant abelian Chern-Simons theory can be written in the form
(\ref{gensolK}), (\ref{gensolT}), this representation is not unique in general. In other words, different $(A,B,C,K,W,t,t',x)$
may correspond to physically equivalent Chern-Simons theories. For this reason, more work is necessary to turn (\ref{gensolK}), 
(\ref{gensolT}) into a one-to-one classification of time reversal invariant abelian insulators. 

\subsection{Examples and physical realizations} \label{examplesec}
In this section, we discuss special cases of the general time reversal
invariant Chern-Simons theories (\ref{gensolK}), (\ref{gensolT}). 

A particularly simple case is when $M=0$ and $W=0$. In this case, (\ref{gensolK}),
(\ref{gensolT}) reduce to
\begin{align}
\mathcal{K} = \bpm K & 0 \\ 0 & -K \epm \ , \ T = \bpm 0 & \bf{1}_N \\ \bf{1}_N & 0 \epm ,
\label{decoupK}
\end{align}
and
\begin{align}
\tau = \bpm t \\ t \epm \ , \ \chi = \bpm 0 \\ t \epm
\label{decoupT}
\end{align}
This theory can be physically realized in a toy model in which spin-up and spin-down electrons
each form completely decoupled quantum Hall states with opposite chiralities: a ``fractional
quantum spin Hall state.'' \cite{BernevigZhang} This model, and in 
particular the stability of the edge modes, was analyzed in Ref. [\onlinecite{LevinStern}]. It is 
also possible to construct models of the type (\ref{decoupK}) from bosons.\cite{Thomale} For example,
we can imagine that bosons of spin $s_z=\pm 1$ form decoupled bosonic quantum Hall states of opposite 
chiralities. In that case $K$ would have only even numbers on the diagonal.

Similarly, one can consider the case where $W = -W^T \neq 0$, so that
\begin{equation}
\mathcal{K} = \bpm K & W \\ W^T & -K \epm
\label{coupK}
\end{equation}
This theory, which was analyzed in Refs. [\onlinecite{Neupertetal}], [\onlinecite{Santosetal}], can also be realized by a toy-model in which spin-up and spin-down electrons form quantum Hall states with opposite chiralities. The main 
difference from the previous case is that here the ground state contains correlations between the two 
spin species. Obviously, the requirement $W = -W^T \neq 0$ necessitates at least a four-dimensional $\mathcal{K}$. 

Next, it is instructive to consider cases where $M=N$, so that 
\begin{align}
\mathcal{K}=\bpm 0 & A \\ A^T & 0 \epm \ , \ T = \bpm -\bf{1}_N & 0 \\ 0 & \bf{1}_N \epm , \ \tau = \bpm 0 \\ t' \epm 
\label{torK}
\end{align}
In this case $\mathcal{K}$ has only even numbers on the diagonal, so that the insulators are bosonic. 
The best known examples of this type have the matrix $A$ being simply an integer number, 
$m$, and $t' = q_B$, where $q_B$ is the boson charge. For $m=1$, this theory describes a conventional 
bosonic Mott insulator. For $m \ne 1$, it describes a charge-conserving model of the 
``toric code''\cite{KitaevToric,LevinBurnellKoch} 
type, with ``charge'' excitations of electric charge $q_B/m$, and ``flux'' excitations which are neutral. 
These examples all have the vector $x=0$ in Eq. \ref{gensolT}, so that $\chi = 0$.

We can also imagine the same class of bosonic insulators, but with $x= 1$. The simplest case, $m=1$ is already
interesting: this theory describes a new kind of bosonic Mott insulator. This insulator can be distinguished from
a conventional Mott insulator by the fact that it has protected edge modes. To see this, consider the edge
stability criterion derived in the next two sections. There, we show that the electronic insulators (\ref{gensolK})
have protected edge modes if and only if $\frac{1}{e^*} \chi^T \mathcal{K}^{-1} \tau$ is odd,
where $e^*$ is the smallest charged excitation of the system in units of $e$. While our derivation focuses on electronic 
insulators, the same analysis applies to bosonic insulators, with the only difference being that we should measure the 
elementary charge $e^*$ in units of the boson charge $q_B$ instead of $e$. Applying this result to the
case $m=1, x=1, t' = q_B$, we find $\frac{1}{e^*} \chi^T \mathcal{K}^{-1} \tau = 1$, implying the existence of
a protected edge mode. Interestingly, this new kind of Mott insulator does not have excitations with fractional
charge or statistics. In this sense, it is similar to a conventional topological insulator, except that it is built out of
bosons instead of fermions. We expect that this ``unfractionalized'' bosonic topological insulator is equivalent
to the one proposed in Ref. [\onlinecite{XieSPT4}]. 

Finally, we discuss examples where all parts of the matrix (\ref{gensolK}) are at play. Recall that
Ref. [\onlinecite{LevinBurnellKoch}] constructed a 2D time reversal invariant lattice model 
built out of electrons described by
\begin{align}
\mathcal{K} = \bpm 0 & m & -k & -k \\
                   m & 0 & 0 & 0 \\
                  -k & 0 & 1 & 0 \\
                  -k & 0 & 0 & -1 \epm \ , \ T = \bpm -1 & 0 & 0 & 0 \\
						       0 & 1 & 0 & 0 \\
						       0 & 0 & 0 & 1 \\
						       0 & 0 & 1 & 0 \epm
\label{exactsolK}
\end{align}
and
\begin{align}
\tau = \bpm 0 \\ 2 \\ 1 \\ 1 \epm \ , \ \chi = \bpm 0 \\ 0 \\ 0 \\ 1 \epm
\label{exactsolT}
\end{align}
In this model the upper-left quadrant of the matrix describes bosons that form a charge-conserving 
toric code\cite{KitaevToric} state, with ``charge'' excitations that carry a charge $2/m$ and ``flux'' 
excitations that are neutral. The off-diagonal quadrant (the $-k$ terms) couple the charge excitations 
with electrons to form composite fermions whose electric charge is $1+2k/m$. The right-bottom quadrant 
then describes how these fermions form a quantum spin Hall state of $\nu=\pm 1$. 

In this example, we can see that the role played by the matrix $B$ in (\ref{gensolK}) is to couple the 
two types of charge excitations to one another. In order to be symmetric to time reversal, this coupling 
must be identical for the two spin directions of the electrons. In a similar fashion, the matrix $C$ in (\ref{gensolK}) 
couples electrons to flux excitations of the bosonic system. In this case, in order for the coupling to preserve 
the symmetry to time reversal, the flux excitations couple oppositely to the two spin directions of the electrons. 

\section{Stability of the edge: microscopic analysis} \label{microedgesec}
An important question is to determine which of the abelian insulators described by (\ref{gensolK}),
(\ref{gensolT}) have protected gapless edge modes -- that is edge modes that cannot be gapped out
without breaking time reversal or charge conservation symmetry, explicitly or spontaneously. 
In the next two sections we show that these systems have protected edge modes if and only if 
the quantity $\frac{1}{e^*}\chi^T \mathcal{K}^{-1} \tau$ is odd. Here, $e^*$ is the smallest charged excitation in the 
system (in units of $e$). Formally, $e^*$ is defined by
\begin{equation}
e^* = \text{min}_l (l^T \mathcal{K}^{-1} \tau)
\label{edef}
\end{equation}
where $l$ ranges over all integer vectors.

This criterion was previously derived in Refs. [\onlinecite{LevinStern}], [\onlinecite{Neupertetal}], and [\onlinecite{LevinBurnellKoch}] 
for the three special cases (\ref{decoupK}), (\ref{coupK}), and (\ref{exactsolK}). In these cases, the criterion can be rephrased in more
physically transparent language, assuming that the system conserves the total electron spin $s^z$. To be specific,
in these cases, we can identify the quantity $\frac{1}{e^*}\chi^T \mathcal{K}^{-1} \tau$ with $-\sigma_{sH}/e^*$
where $\sigma_{sH}$ is the spin-Hall conductivity in units of $e/2\pi$. Thus, the criterion reduces to the statement
that these systems have protected edge modes if and only if $\sigma_{sH}/e^*$ is odd. \cite{LevinStern}
More generally, this reformulation in terms of $\sigma_{sH}/e^*$ is valid for any $s^z$ conserving insulator for which
$\chi \equiv \tau_\down \text{ (mod $2$)}$, where $\tau_\down$ is the charge vector corresponding to the
spin-down electrons (see appendix \ref{shcritapp}). Physically, the condition $\chi \equiv \tau_\down \text{ (mod $2$)}$ 
is equivalent to the requirement that the excitations of the insulator have a ``local Kramers degeneracy'' (section \ref{lockramsec}) 
if and only if they carry half-integer spin. This connection between spin and time reversal is what allows us to reformulate 
the edge stability criterion in terms of $\sigma_{sH}$.

\subsection{Basic setup}
We derive the $\frac{1}{e^*}\chi^T \mathcal{K}^{-1} \tau$ criterion in two steps. First, we
use a microscopic approach to explicitly show that when $\frac{1}{e^*}\chi^T \mathcal{K}^{-1} \tau$ is even, the 
edge can be gapped out without breaking any symmetries (explicitly or spontaneously). We also show that there 
is an obstruction to gapping out the edge when $\frac{1}{e^*}\chi^T \mathcal{K}^{-1} \tau$ is odd. Then, in section 
\ref{fluxargsec}, we complete the derivation by giving a general argument that the edge is protected when
$\frac{1}{e^*}\chi^T \mathcal{K}^{-1} \tau$ is odd.

Our analysis closely follows that of Ref. [\onlinecite{LevinStern}].
Our starting point is the edge theory (\ref{kedge}), which we reprint below for convenience:
\begin{eqnarray*}
L &=& \frac{1}{4\pi} (\mathcal{K}_{IJ} \partial_x \Phi_I \partial_t \Phi_J
- \mathcal{V}_{IJ} \partial_x \Phi_I \partial_x \Phi_J) \nonumber \\
&+& \frac{1}{2\pi} \epsilon^{\mu \nu} \tau_I \partial_\mu \Phi_I A_\nu
\end{eqnarray*}
We recall that operators of the form $e^{i\Theta(\Lambda)}$, with $\Theta(\Lambda) \equiv \Lambda^T \mathcal{K} \Phi$, 
are local in the sense that they correspond to products of electron creation and annihilation operators.

This edge theory has $2N$ gapless edge modes, $N$
for each chirality. Our goal is to find the conditions under
which these modes can be gapped out by charge conserving, time
reversal symmetric perturbations. We focus on a general class of scattering terms of the form
\begin{equation}
U(x)\cos(\Theta(\Lambda) - \alpha(x))
\end{equation}
where $\Lambda$ is a $2N$-dimensional integer valued vector.
Imposing charge conservation leads to the requirement
$\Lambda^T \tau = 0$. As for time reversal, we note that Eq. \ref{gentedge}
implies that $\Theta$ transforms under time reversal as
\begin{equation}
\mathcal{T}\Theta(\Lambda) \mathcal{T}^{-1}
=\Theta(-T\Lambda)-Q(\Lambda)\pi,
\label{timereversal}
\end{equation}
where
\begin{equation}
Q(\Lambda)\equiv -\Lambda^T\chi
\end{equation}
Thus, one can construct scattering terms that are even/odd under time reversal by
defining
\begin{eqnarray}
U_\pm(\Lambda) &=& U(x) [\cos(\Theta(\Lambda) - \alpha(x)) \nonumber \\
    &\pm& (-1)^{Q(\Lambda)}\cos(\Theta(-T\Lambda) -\alpha(x)) ]
\label{pert3}
\end{eqnarray}
In general, we should include Klein factors in the definition of $U_{\pm}(\Lambda)$ to ensure that these
terms obey the correct commutation relations -- i.e., these terms should commute with each other 
when they act at spatially separated points. However, in the analysis below, we only consider sets of
$\{\Lambda_i\}$ satisfying $\Lambda_i^T \mathcal{K} \Lambda_j = 0$. This condition guarantees that 
the $e^{i\Theta(\Lambda)}$ operators automatically commute with one another, without any need for Klein factors. 
\footnote{It is not hard to show that $e^{i\Theta(\Lambda)}$ and $e^{i\Theta(\Lambda')}$
commute if $\Lambda^T \mathcal{K} \Lambda'$ is even and anti-commute if $\Lambda^T \mathcal{K} \Lambda'$ is odd, using 
$[\phi_I(x), \phi_J(y)] = \pi i \mathcal{K}^{-1}_{IJ} \text{sgn}(x-y)$.} Thus, the Klein factors 
can be safely ignored, and we will drop them for clarity. 

We now examine whether time reversal symmetric terms of the type $U_+$ (\ref{pert3}) can gap the spectrum without spontaneously
breaking time reversal symmetry. Importantly, it will \emph{not} matter to us whether these terms are relevant or irrelevant in the
renormalization group sense. One reason is that we are 
interested in whether any term can gap out the edge, including those with large coefficients. Thus, the perturbative stability of
the edge is not our concern here. Another reason is that we can always make the $\{U_+(\Lambda)\}$ terms relevant by appropriately tuning 
$\mathcal{V}_{IJ}$ -- at least when $\{\Lambda_i\}$ satisfy $\Lambda_i^T \mathcal{K} \Lambda_j = 0$, as 
assumed below.

Before tackling the general case, we first warm up with two simple examples.
Both examples were discussed (briefly) in Ref. [\onlinecite{LevinStern}].

\subsection{Example 1: Laughlin quantum spin Hall state}
In this section, we analyze the stability of the edge for the case
\begin{align}
\mathcal{K} = \bpm k & 0 \\
		   0 & -k \epm , \ T = \bpm 0 & 1 \\ 1 & 0 \epm , \ \tau = \bpm 1 \\ 1 \epm , \  \chi = \bpm 0 \\ 1 \epm
\end{align}
Physically, this case can realized by a toy model in which the spin-up and spin-down electrons
each form $\nu = 1/k$ Laughlin states with opposite chiralities.

We note that the elementary charge is
\begin{equation}
e^* = \text{min}_l(l^T \mathcal{K}^{-1} \tau) = \frac{1}{k}
\end{equation}
while the quantity $\chi^T \mathcal{K}^{-1} \tau$ is given by
\begin{equation}
\chi^T \mathcal{K}^{-1} \tau = -\frac{1}{k}
\end{equation}
In particular, we have $\frac{1}{e^*}\chi^T \mathcal{K}^{-1} \tau=-1$ for all $k$. Thus, according to the
general criterion, all of these states have protected edge modes.

To see how this stability manifests itself in a microscopic analysis, note
that the only charge conserving vectors are of the form $\Lambda=(n,-n)$.
The corresponding perturbation
$U(x) \cos(\Theta(n,-n) - \alpha(x))$ is even
under time reversal for even $n$ and odd for odd $n$. Thus,
time reversal symmetry requires even $n$, say $n=2$.
Adding such a perturbation will indeed open a gap in the spectrum --
at least for large $U$. However, hand in hand with gapping out the spectrum, 
this perturbation also spontaneously breaks time reversal symmetry. To see this, note that when 
the perturbation gaps out the edge, it ``freezes'' the value of $\Theta(n,-n)$. 
As a result, it also freezes the value of $\Theta(1,-1)$. But then $\cos(\Theta(1,-1) - \alpha)$ -- an operator which
is odd under time reversal -- acquires a non-zero expectation value. It follows that time reversal symmetry is broken
spontaneously. 
In this way, we see that none of these perturbations can gap the two edge modes without breaking time
reversal symmetry, either explicitly or spontaneously.

\subsection{Example 2: Two component fractional quantum spin Hall state} \label{twocompsec}
In this section, we analyze the stability of the edge for the case
\begin{align}
\mathcal{K} = \bpm K & 0 \\ 0 & -K \epm \ , \ T = \bpm 0 & \bf{1} \\ \bf{1} & 0 \epm , \nonumber \\ 
\tau^T = (1 , 1 , 1 , 1 ) \ , \  \chi^T = (0, 0, 1, 1)
\end{align}
where $K$ is a $2 \times 2$ matrix. Physically, this case can be realized
by a toy model in which the spin-up and spin-down electrons each form two 
component quantum Hall states with opposite chiralities.

In this case, there are two pairs of counter-propagating edge modes, so we need to fix
the value of two different $\Theta(\Lambda)$'s to gap out the
edge. The simplest way to do this is to add a perturbation of the form
$U_+(\Lambda)$ (\ref{pert3}) where $\Lambda, -T \Lambda$ are linearly
independent and charge conserving. We write this term as
\begin{eqnarray}
&U(x)& [\cos(\Theta(\Lambda_1) - \alpha(x)) \nonumber \\
&+& (-1)^{Q(\Lambda_1)} \cos(\Theta(\Lambda_2) - \alpha(x))]
\label{pert4}
\end{eqnarray}
where $\Lambda_1 \equiv \Lambda$ and $\Lambda_2 \equiv -T\Lambda$. 

According to Haldane's null vector criterion\cite{Haldane},
such a perturbation can gap out the $4$ edge modes if $\Lambda_1, \Lambda_2$ satisfy
\begin{equation}
\Lambda_1^T \mathcal{K} \Lambda_1 = \Lambda_2^T \mathcal{K}
\Lambda_2 = \Lambda_1^T \mathcal{K} \Lambda_2 = 0
\label{cond1}
\end{equation}
The origin of this criterion is that it guarantees that one can make a linear change of variables from
$\Phi$ to $\Phi'$ such that the action for $\Phi'$ will be that of two
decoupled non-chiral Luttinger liquids. The two terms in (\ref{pert3}) will then
gap the spectrum of these two liquids by freezing the values of $\Theta(\Lambda_1)$
and $\Theta(\Lambda_2)$. We review this result in appendix \ref{nullvecapp}.

We now turn to search for charge conserving $\Lambda_1$ and $\Lambda_2$ such that
$\Lambda_2 = -T \Lambda_1$, and such that $\Lambda_1, \Lambda_2$ satisfy the condition (\ref{cond1}). It is convenient to
parameterize the matrix $\mathcal{K}$ as
\begin{equation}
\mathcal{K} = \bpm b+us & b & 0 & 0 \\
		    b   & b+vs & 0 & 0 \\
		    0 & 0 & -b-us & -b \\
		    0 & 0 & -b & -b-vs \epm
\end{equation}
with $b,u,v,s$ integers and $u$ and $v$ having no common factor. In terms of these
parameters, the elementary charge is
\begin{equation}
e^* = \text{min}_l(l^T \mathcal{K}^{-1} \tau) = \frac{1}{(u+v)b+uvs}
\end{equation}
Also, the quantity $\chi^T \mathcal{K}^{-1} \tau$ is given by
\begin{equation}
\chi^T \mathcal{K}^{-1} \tau = - \frac{u+v}{(u+v)b+uvs}
\end{equation}
The ratio $\frac{1}{e^*} \chi^T \mathcal{K}^{-1} \tau$ is then $(u+v)$, so according to the general criterion,
the parity of $u+v$ determines whether the spectrum can be gapped.

When $u+v$ is odd, it is indeed impossible to find $\Lambda_1,
\Lambda_2$ that do not spontaneously break time
reversal symmetry. Imagine one had such
a solution and define $\Lambda_{\pm} = \Lambda_1 \pm \Lambda_2$. Then,
\begin{align}
T \Lambda_{-} = \Lambda_{-} \ , \ Q(\Lambda_{-}) = 0
\end{align}
so $\Lambda^T_{-}$ must be an integer multiple of $(1,-1,1,-1)$. Also,
\begin{align}
T \Lambda_{+} = -\Lambda_{+} \ , \ \Lambda_{-}^T \mathcal{K}\Lambda_+ = 0
\label{lambdaplus}
\end{align}
so $\Lambda^T_+$ must be an integer multiple of
$(v,u,-v,-u)$. But $\cos(\Theta(v,u,-v,-u)-\alpha)$ is odd under time reversal,
according to (\ref{timereversal}).
This means that the perturbation (\ref{pert4}) will spontaneously break time reversal
symmetry: the scattering term (\ref{pert4}) freezes the value of $\Theta(\Lambda_1)$,
$\Theta(\Lambda_2)$ and therefore also freezes the value of $\Theta(\Lambda_1)+\Theta(\Lambda_2)$
and hence $\Theta(v,u,-v,-u)$. It follows that $\cos(\Theta(v,u,-v,-u)-\alpha)$ acquires
an expectation value, spontaneously breaking time reversal symmetry.

On the other hand, when $u+v$ is even (so that both $u,v$ are
odd), the above analysis suggests an obvious solution $(\Lambda_1,
\Lambda_2)$.  We can take
\begin{eqnarray}
\Lambda^T_{-} &=& (1,-1,1,-1) \\
\Lambda^T_{+} &=& (v,u,-v,-u)
\end{eqnarray}
so that
\begin{eqnarray}
\Lambda^T_1 &=& \frac{1}{2}\left(1+v,-1+u, 1-v, -1-u \right) \nonumber \\
\Lambda^T_2 &=& \frac{1}{2}\left(-1+v, 1+u, -1-v, 1-u \right)
\label{evennumv}
\end{eqnarray}

To complete the analysis, we need to check that the perturbation
(\ref{pert4}) corresponding to $\Lambda_1, \Lambda_2$ does not spontaneously
break time reversal symmetry. The only
way that time reversal symmetry (or any other symmetry) can be
spontaneously broken is if, for some $a_1, a_2$ with no common
factors, the linear combination $a_1 \Lambda_1 + a_2 \Lambda_2$
is non-primitive -- that is, $a_1 \Lambda_1 + a_2 \Lambda_2 =
k \Lambda$ where $\Lambda$ is an integer vector, and $k$
is an integer larger than $1$. If such an $a_1, a_2, \Lambda$ exist,
then when the perturbation freezes the value of $\Theta(\Lambda_i)$
it will also freeze the value of $\Theta(\Lambda)$
which may lead to spontaneous symmetry breaking,
as in the discussion after (\ref{lambdaplus}). Conversely, if all
such linear combinations are primitive, then the perturbation
does not break any symmetries when it freezes $\Theta(\Lambda_i)$.
\emph{Thus, we can be assured that no symmetries are broken spontaneously,
as long as we choose the $\Lambda$'s so that all linear
combinations $a_1 \Lambda_1 + a_2 \Lambda_2$ are primitive}.

It is possible to show that such a non-primitive $a_1 \Lambda_1 + a_2 \Lambda_2$
exists if and only if the six $2 \times 2$ minors generated from the
$4 \times 2$ matrix with columns $\Lambda_1, \Lambda_2$ have a common
factor (see appendix \ref{primcondapp}). We now check this condition explicitly. Writing out the $4 \times 2$ matrix 
corresponding to (\ref{evennumv}) gives
\begin{equation}
\frac{1}{2} \cdot\bpm 1+v & -1+v \\
    		     -1+u & 1+u \\
                      1-v & -1-v \\
                     -1-u & 1-u \epm,
\end{equation}
We can see that the minor corresponding to the first two rows is given by
\begin{equation}
\frac{1}{4} \cdot [(1+v)(1+u) - (-1+v)(-1+u)] = \frac{u+v}{2}
\label{minorex1}
\end{equation}
while the minor corresponding to the second and third rows is
\begin{equation}
\frac{1}{4} \cdot [(-1+u)(-1-v) - (1+u)(1-v)] = \frac{v-u}{2}
\label{minorex2}
\end{equation}
Clearly these two minors have no common factor since $u,v$
have no common factor. We conclude that there are no non-primitive linear combinations 
$a_1 \Lambda_1 + a_2 \Lambda_2$ and therefore the perturbation does not break time 
reversal symmetry (or any other symmetry) spontaneously.

\subsection{Microscopic analysis in general case}
In this section, we analyze the edge stability of the general abelian insulators described by
(\ref{gensolK}), (\ref{gensolT}). The simplest way to gap the edge in the general case is to add several 
perturbations of the form $U_+(\Lambda)$ (\ref{pert3}) each
with a different $\Lambda$. In general, these perturbations can be divided into two different
classes: perturbations where $\Lambda, -T \Lambda$ are linearly independent and perturbations where
$T \Lambda = \pm \Lambda$. Perturbations of the first type can gap out two pairs of edge modes while
perturbations of the second type can gap out a single pair of edge modes. Therefore, in order to gap
out all $N$ pairs of edge modes, we need $k$ perturbations of the first type and $l$ perturbations of
the second type, where $2k+l = N$.

We can describe the sum of all these perturbations by listing all the linearly independent $\Lambda$'s
that appear in the various cosine terms. All together, we will have $2k+l = N$ different $\Lambda$'s since
the first type of perturbation involves two $\Lambda$'s (namely $\Lambda, -T \Lambda$) while the
second type of perturbation involves one $\Lambda$. We will label the $\Lambda$'s by $\Lambda_1,...,\Lambda_N$.

Just like the example discussed in the previous section, the perturbations corresponding to $\Lambda_1,...,\Lambda_N$
can gap the edge if the $\{\Lambda_i\}$ satisfy Haldane's null vector condition\cite{Haldane}
\begin{equation}
\Lambda_i^T \mathcal{K} \Lambda_j = 0
\label{gennull}
\end{equation}
This condition guarantees that we can make a change of variables from $\Phi$ to $\Phi'$ such that (a) the
resulting action consists of $N$ decoupled non-chiral Luttinger liquids, and (b) the cosine terms
correspond to backscattering terms for each of these decoupled liquids. (For an example, see
appendix \ref{nullvecapp}).

We now turn to search for charge conserving $\{\Lambda_i\}$ satisfying
these conditions. It is convenient to define a vector
\begin{equation}
\Lambda_c = \frac{1}{e^*}\mathcal{K}^{-1} \tau
\label{lambdacdef}
\end{equation}
We note that the definition of $e^*$ (\ref{edef}) implies that
$\Lambda_c$ is an integer vector and that the greatest common divisor
of $\{\Lambda_{ci}\}$ is $1$. According to our criterion, 
the edge is protected if and only if the quantity $\chi^T \Lambda_c$ is 
odd. We now verify this claim.

When $\chi^T \Lambda_c$ is odd, it is indeed impossible to find
$\{\Lambda_i\}$ which satisfy (\ref{gennull}) and which do not spontaneously break time reversal
symmetry. The reason is that one can always find a linear combination of $\Lambda_i$
which is a multiple of $\Lambda_c$, as we prove in the next paragraph. But $\cos(\Theta(\Lambda_c)-\alpha)$
is odd under time reversal according to (\ref{timereversal}). It then follows that the perturbations
corresponding to $\{\Lambda_i\}$ will spontaneously break time reversal symmetry:
these perturbations will freeze the values of $\Theta(\Lambda_i)$ and therefore
also $\Theta(\Lambda_c)$, thus giving an expectation value to the time reversal
odd operator $\cos(\Theta(\Lambda_c)-\alpha)$.

To complete the argument, we explain why one can always find a linear combination of
$\{\Lambda_i\}$ which is a multiple of $\Lambda_c$. First, we note that $\Lambda_i^T \mathcal{K} \Lambda_c = 0$
for all $i$, by charge conservation. At the same time, we can see that the set of all vectors
satisfying $\{\Lambda_i^T \mathcal{K} \Lambda = 0\}$ has dimension $N$, since these relations describe $N$ equations
in $2N$ unknowns. Combining this observation with the fact that the $\{\Lambda_i\}$ themselves provide $N$
linearly independent solutions to these equations, we deduce that $\Lambda_c$ cannot be linearly
independent from $\{\Lambda_i\}$. Hence, there must be a linear relation of the form
$a_c \Lambda_c + \sum a_i \Lambda_i = 0$ with all the $a$'s being integers. Intuitively, the basic point of this 
argument is that one cannot gap the edge without gapping the ``charge mode'' $\Theta(\Lambda_c)$, and this mode is protected by
time reversal symmetry whenever $\chi^T \Lambda_c$ is odd.

On the other hand, if $\Lambda_c^T \chi$ is even, then the edge \emph{can} be gapped out,
as we now show. It is convenient to work in a basis where $t = (1,1,...,1)$ and
$t' = (2,2,...,2)$. Using the fact that $T \Lambda_c = - \Lambda_c$, we can see that $\Lambda_c$ is of the form
\begin{equation}
\Lambda_c^T = (w, 0_M, u, -u)
\end{equation}
where $w = (w_1, ..., w_M)$ is an $M$ component integer vector,
$u = (u_1, ..., u_{N-M})$ is an $N-M$ component integer vector and
$0_M$ denotes an $M$ component vector of $0$'s. 

It is convenient to divide our analysis into two cases: either
(a) the vector $w$ has at least one odd entry or (b) the vector $w$ only has even entries.
We begin with case (a). First, we introduce some notation.
We define $N-M$ vectors $e_1, ..., e_{N-M}$, each with $N-M$ components, by
\begin{eqnarray}
e_1 &=& (1, 0, 0, ..., 0) \\
e_2 &=& (0, 1, 0, ..., 0) \nonumber \\
e_3 &=& (0, 0, 1, ..., 0) \nonumber \\
&\vdots& \nonumber
\end{eqnarray}
and $M$ vectors $f_1, ..., f_{M}$, each with $M$ components, by
\begin{eqnarray}
f_1 &=& (1, 0, 0, ..., 0) \\
f_2 &=& (1, -1, 0, ..., 0) \nonumber \\
f_3 &=& (1, 0, -1, ..., 0) \nonumber \\
&\vdots& \nonumber
\end{eqnarray}
We then define $N-1$ vectors $\Xi_1, ..., \Xi_{N-1}$, each with $2N$ components by
\begin{eqnarray}
\Xi_i &=& (0_M, f_{i+1}, 0_{N-M}, 0_{N-M}) ,\ 1 \leq i \leq M-1  \label{xi} \\
\Xi_i &=& (0_M, f_1, -e_{i-M+1}, -e_{i-M+1}) ,\  M \leq i \leq N-1 \nonumber
\end{eqnarray}
These vectors obey the conditions $T \Xi_i = \Xi_i$,
and $\Xi_i^T \mathcal{K} \Xi_j = 0$. Furthermore, since these vectors are
charge conserving, we have $\Xi_i^T \mathcal{K} \Lambda_c = \frac{1}{e^*} \Xi^T \tau = 0$.
Therefore, if we define $\Lambda_1 = \Lambda_c, \Lambda_2 = \Xi_1, \Lambda_3 = \Xi_2$, etc.,
we will have a set of $N$ vectors satisfying condition (\ref{gennull}). 

To complete the argument we need to check that these perturbations
do not spontaneously break time reversal symmetry. Like the example discussed
in the previous section, it is sufficient to show that for any set of $a_i$ which have no common factor,
the linear combination $\sum_i a_i \Lambda_i$ is always primitive. This is turn reduces
to the question of showing that the $\bpm 2N \\ N \epm$ $N \times N$
minors of the matrix with columns $\Lambda_i$ have no common factors (see appendix \ref{primcondapp}).
This property can be established straightforwardly, using the
fact that the greatest common divisor of $\{\Lambda_{ci}\}$ is $1$. For a simple
example, see Eqs. \ref{minorex1}-\ref{minorex2}.

We now consider case (b): the vector $w$ has only even entries. In this case, we define a vector
\begin{equation}
\Lambda_n^T = (0_M, \alpha_{tot} \cdot f_1, -\alpha, -\alpha)
\end{equation}
where $\alpha = (\alpha_1, ..., \alpha_{N-M})$, $\alpha_{tot} = \sum_i \alpha_i$,
and $\alpha_i$ is $0$ or $1$ depending on whether $u_i$ (\ref{lambdacdef}) is even or odd.

We then let
\begin{align}
\Lambda_{1} = \frac{1}{2}(\Lambda_c + \Lambda_n) \ , \ \Lambda_{2} = \frac{1}{2}(\Lambda_c - \Lambda_n) 
\label{lambda12}
\end{align}
Just as before, one can check that these two vectors
obey the condition
$\Lambda_1^T \mathcal{K} \Lambda_1 = \Lambda_2^T \mathcal{K}
\Lambda_2 = \Lambda_1^T \mathcal{K} \Lambda_2 = 0$.
Thus, we now have $2$ vectors satisfying condition (\ref{gennull}).

To fully gap out the edge, we need to find $N-2$ more vectors.
These can be obtained as follows. Consider the $N-1$ vectors
$\Xi_i$ (\ref{xi}). These vectors obey the conditions $T \Xi_i = \Xi_i$,
and $\Xi_i^T \mathcal{K} \Xi_j = 0$. Furthermore, by the same reasoning as 
above, we have $\Xi_i^T \mathcal{K} \Lambda_1 = \Xi_i^T \mathcal{K} \Lambda_2 = 0$.
Therefore, if we define $\Lambda_3 = \Xi_1, \Lambda_4 = \Xi_2, ...$
we will have a set of $N+1$ vectors satisfying condition (\ref{gennull}).
The only problem is that these vectors are not
all linearly independent since $\Lambda_n = \sum_i \alpha_i \Xi_{i+M-1}$.
Hence, we need to drop an appropriate one of the $\Xi_{i}$'s. Once we do this,
we will have the required set of $N$ vectors. As in case (a), we can
verify that these perturbations do not spontaneously break time reversal
symmetry by showing that the $\bpm 2N \\ N \epm$ $N \times N$
minors of the matrix with columns $\Lambda_i$ have no common factors.

Strictly speaking, we are not quite finished, since we implicitly
assumed that $M \neq 0$. If $M=0$, our construction of the scattering
terms $\Lambda_i$ needs to be slightly modified. In this case, we define
\begin{equation}
\Lambda_n^T = (\sum_{i=2}^{N} \alpha_i, -\alpha_2,...,-\alpha_{N}, \sum_{i=2}^{N-1} \alpha_i, -\alpha_2,...,-\alpha_{N})
\end{equation}
Also, we define $\Xi_1, ... \Xi_{N-1}$ by
\begin{eqnarray}
\Xi_1 &=& (1, -1, 0, 0, ..., 1, -1, 0, 0, ...) \\
\Xi_2 &=& (1, 0, -1, 0, ..., 1, 0, -1, 0, ...) \nonumber \\
\Xi_3 &=& (1, 0, 0, -1, ..., 1, 0, 0, -1, ...) \nonumber \\
\vdots \nonumber
\end{eqnarray}
We then define $\Lambda_1, \Lambda_2$ as in (\ref{lambda12}) 
and take $\Lambda_3 = \Xi_1, \Lambda_4 =  \Xi_2, ...$ as above.
The remainder of the analysis is identical to case (b), studied above.

\section{Stability of the edge modes: flux insertion argument} \label{fluxargsec}
In the previous section, we found that when $\frac{1}{e^*}\chi^T \mathcal{K}^{-1} \tau$
was odd, there was no simple way to gap out the edge modes without breaking time reversal
or charge conservation symmetry. However, that analysis does not rule out
the possibility of gapping out the edge using other, more complicated,
perturbations. In this section, we fill in this hole: we give a general argument proving
that it is impossible to gap the edge when $\frac{1}{e^*}\chi^T \mathcal{K}^{-1} \tau$ 
is odd. Combining this result with the conclusions of section
\ref{microedgesec} completes our proof that abelian time reversal invariant insulators have
protected gapless edge modes if and only if the quantity $\frac{1}{e^*}\chi^T \mathcal{K}^{-1} \tau$ is odd.

The argument we present is very similar to
the one given in Ref. [\onlinecite{LevinStern}] which is in turn a generalization of the
flux insertion argument of Ref. [\onlinecite{FuKanePump}]. The statement we prove is as
follows: we consider one of the insulators (\ref{gensolK}), (\ref{gensolT}) 
in a cylindrical geometry with an even number of electrons and zero flux through the cylinder. 
We also assume that the ground state is time reversal invariant and has short range
correlations. Assuming that the quantity $\frac{1}{e^*}\chi^T \mathcal{K}^{-1} \tau$ is odd, 
we prove that the system always contains at least one 
low lying excited state -- that is, an excited state whose energy gap 
vanishes in the thermodynamic limit. Furthermore, this excited state has the important property that
it is in the same ``topological sector'' as the ground state (as will be explained below). We interpret 
this low lying state as evidence for a protected gapless edge mode.

%\begin{figure}[h]
\begin{figure}[tb]
\centerline{
\includegraphics[width=0.8\columnwidth]{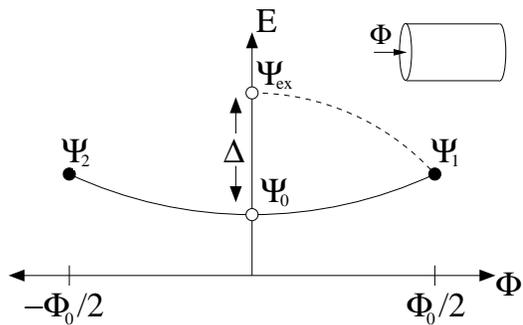}
}
\caption{
The flux insertion argument for the $\nu = k$ toy model: we start with the ground state
$\Psi_0$ and adiabatically inset $\pm \Phi_0/2$ flux through the cylinder,
obtaining two states $\Psi_1, \Psi_2$. If $k$ is odd, then $\Psi_1$ has a
Kramers degeneracy at the two ends of the cylinder, and is therefore degenerate
in energy with three other states, one of which is $\Psi_2$. If we start in one
of these three degenerate states and then adiabatically reduce the flux to $0$,
we obtain an excited state $\Psi_{ex}$ whose energy gap $\Delta$ vanishes in the
thermodynamic limit.
}
\label{adiabatic}
\end{figure}

\subsection{Integer case}
We begin by explaining the argument in a simple case. To be specific, we consider a non-interacting
toy model where spin-up and spin-down electrons each form $\nu = k$ integer quantum Hall
states with opposite chiralities. (This example corresponds to the quantum spin Hall 
state (\ref{decoupK}) with $K = \bf{1}_k$). As mentioned above, we consider this model in a cylindrical
geometry with an even number of electrons and zero flux through the cylinder.
We assume that the ground state is time reversal invariant and has short range correlations. We will show that if $k$ is odd 
there is always at least one low lying excited state -- that is a state whose energy gap vanishes in the
thermodynamic limit. Furthermore, this state is robust if we add arbitrary time reversal invariant, 
charge conserving local perturbations to the Hamiltonian, as long as we do not close the bulk gap
or spontaneously break one of the symmetries.

To begin, we consider the ground state $|\Psi_0\>$ of the toy model at zero flux and
imagine adiabatically inserting $\Phi_0 /2 = hc/2e$ flux through the cylinder.
Let us call the resulting state $|\Psi_1\>$. Similarly, we let $|\Psi_2\>$ be the state obtained
by adiabatically inserting $-\Phi_0/2$ flux (see Fig. \ref{adiabatic}). We note that $|\Psi_1\>, |\Psi_2\>$
are time reversed partners: $|\Psi_2\> = \mathcal{T} |\Psi_1\>$. The remainder of the argument can be divided
into two parts. In the first part, we show that when $k$ is odd, $|\Psi_1\>$ has a Kramers 
degeneracy at the two ends of the cylinder (a precise definition of this notion of ``local Kramers degeneracy''
is given in section \ref{lockramsec}). In the second part, we use this Kramers degeneracy
to prove that there is a robust low lying excited state at zero flux.

\begin{figure}[tb]
\centerline{
\includegraphics[width=0.7\columnwidth]{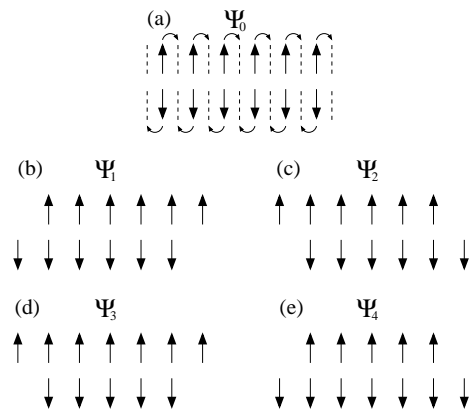}
}
\caption{
(a) A schematic portrait of the ground state $\Psi_0$ of the $\nu = 1$ toy
model. Working in the Landau gauge, the single particle states consist of spin-up and
spin-down Landau orbitals. In the time reversal invariant ground state, the spin-up and spin-down
orbitals are equally occupied. After inserting $\Phi_0/2$ flux, the spin-up and spin-down
orbitals shift in opposite directions, resulting in the state $\Psi_1$. (b) The state
$\Psi_1$ has a Kramers degeneracy on both ends of the cylinder and is therefore degenerate
with its time reversed partner (c) $\Psi_2$, as well as two other states, (d) $\Psi_3$ and
(e) $\Psi_4$.
}
\label{psi}
\end{figure}

We begin with the first part -- establishing that $|\Psi_1\>$ has a Kramers degeneracy
at the two ends of the cylinder. Here, we give an intuitive argument; we derive this claim
rigorously (and in greater generality) in section \ref{lockramsec}. It is useful to start with the
$\nu = 1$ case. Consider the $\nu = 1$ toy model in the Landau gauge. The single particle
eigenstates for spin-up and spin-down electrons are then Landau
orbitals, localized in the longitudinal direction and delocalized in the periodic direction.
Each spin-up orbital has a corresponding time-reversed spin-down partner. These pairs are
equally occupied with spin-up and spin-down electrons in the ground state $|\Psi_0\>$, since by
assumption, $|\Psi_0\>$ is time reversal invariant (see Fig. \ref{psi}a). If we now adiabatically insert
$\Phi_0/2$ flux, each spin-up orbital shifts to the right by half of the inter-orbital spacing,
while each spin-down orbital shifts to the left. The resulting state $|\Psi_1\>$ has one unpaired
spin-up electron on the right edge and one unpaired spin-down electron on the left edge (Fig. \ref{psi}b).
More generally, for the $\nu = k$ toy model, one finds that $|\Psi_1\>$ has $k$ unpaired spins on the two edges.
We can now see that when $k$ is odd, $|\Psi_1\>$ has an odd number of electrons localized near the two edges.
On an intuitive level, this property implies our result: $|\Psi_1\>$ has a Kramers
degeneracy at each of the two ends of the cylinder.

We now explain the second part of the argument -- why the Kramers degeneracy
at half a flux quantum implies that there is a robust
low-lying excited state at zero flux. The basic point is very simple: as long as the
two ends of the cylinder are well separated, Kramers theorem guarantees that
$|\Psi_1\>$ is part of a multiplet of $4$ states which are nearly degenerate in energy
(see Fig. \ref{psi}b-e). We note that $|\Psi_2\>$ is one of these degenerate states, as it is
the time reversed partner of $|\Psi_1\>$. We now imagine starting at $\Phi_0/2$ flux and then
adiabatically reducing the flux to $0$.
If we start with the state $|\Psi_1\>$, then adiabatic flux removal takes us to
the ground state $|\Psi_0\>$. However, if we start with $|\Psi_2\>$, or one of the other $2$ states
degenerate with $|\Psi_1\>$, the result is an eigenstate
$|\Psi_{ex}\>$ of the zero flux Hamiltonian which is distinct from $|\Psi_0\>$
(see Fig. \ref{adiabatic}). At the same time, it necessarily has low energy since the energy
change $\Delta$ associated with an adiabatic insertion of flux through a
cylinder must vanish in the thermodynamic limit (assuming charge conservation is not
broken spontaneously). In this way, we can construct a low-lying excited state $|\Psi_{ex}\>$ 
whose energy gap vanishes in the thermodynamic limit.

To complete the argument, we now imagine adding an arbitrary local, time
reversal invariant, charge conserving perturbation to the system
(for example, we could add short-ranged interactions between the electrons).
As long as the perturbation does not close the bulk gap, the above picture must stay the
same: the local Kramers degeneracy between $|\Psi_1\>$, $|\Psi_2\>$ must remain intact, 
and hence $|\Psi_{ex}\>$ must continue to be low in energy. We conclude that the system always contains at least
one low-lying excited state $|\Psi_{ex}\>$ whose energy gap vanishes in
the thermodynamic limit.

\subsection{Topological order and ground state degeneracy on a cylinder}
One of the main complications in extending the flux insertion argument to the
general case is that the more general states (\ref{gensolK}), (\ref{gensolT}) have
``topological order''\cite{WenReview,WenBook} -- that is, they support quasiparticle excitations
with fractional statistics. An important consequence of this is that these systems typically have
multiple degenerate ground states when defined in a topologically non-trivial geometry,
such as a torus. \cite{WenReview,WenBook,Einarsson} This ground state degeneracy is very robust
and cannot be split by any local perturbation (in the thermodynamic limit).

Similar low-lying states can occur in a cylindrical geometry -- the geometry of interest here.\cite{GefenThouless}
As an example, consider a toy model where spin-up and spin-down electrons each form $\nu = 1/3$
Laughlin states with opposite chiralities (this model corresponds to the case (\ref{decoupK}) with $K=3$). 
This model has three protected low-lying states in a cylindrical
geometry. The first low-lying state is the ground state $|\Psi_0\>$. The other two states can be obtained by
starting with $|\Psi_0\>$ and adiabatically inserting either $\Phi_0$ flux, yielding a state
$|\Psi_0'\>$, or $2\Phi_0$ flux, yielding $|\Psi_0''\>$. We know that these states are (nearly) degenerate
in energy, since the energy change $\Delta$ associated with inserting a fixed amount of flux must vanish in
the thermodynamic limit (assuming that charge conservation is not broken). At the same time, we can
see that they are orthogonal to one another since the insertion of a flux quantum transfers
a spin-up $e/3$ quasiparticle from the left edge to the right edge, and a spin-down $e/3$ quasiparticle from the
right edge to the left edge. Given that the states differ in the amount of spin on the two edges, they must be orthogonal.

In fact, the three states are not only orthogonal, but they belong to different ``topological sectors.''
That is, they cannot be coupled together by any $\mathcal{O}$ which is a finite product of local operators:
\begin{equation}
\<\Psi_0 | \mathcal{O}|\Psi_0'\> = \<\Psi_0 | \mathcal{O}|\Psi_0''\> = \<\Psi_0' | \mathcal{O}|\Psi_0''\> = 0
\end{equation}
On an intuitive level, these matrix elements vanish because an $e/3$ quasiparticle is a fractionalized excitation 
with nontrivial statistics, so no local operator composed out of electron creation and annihilation operators can 
create or destroy such an excitation. (Instead, a ``string-like'' operator is required to move such an excitation from 
one edge to another). A more precise way to argue this is to note that if one takes a pair of states, say 
$|\Psi_0\>, |\Psi_0'\>$, and measures the Berry phase associated with moving an $e/3$ spin-up quasiparticle around 
the cylinder, one gets a different result for the two states (with the difference given by the statistical phase 
$2\theta_{ex} = 2\pi/3$). No local operator (or finite product of local operators) can change this relative Berry phase and hence no 
term of this kind can connect the two states.

\subsection{General case: outline of the argument} \label{genfluxargsec}
As explained above, when a topologically ordered system is defined in a cylindrical geometry,
one typically finds multiple low-lying states -- each one belonging to a different topological
sector. Because of this phenomenon, we will raise our standards for the flux
insertion argument. It is not enough to just show that there are
low-lying states at zero flux; this is (nearly) always the case in topologically ordered systems and
does not constitute evidence for a protected gapless edge mode, even at a heuristic level.
Instead we will show that there is a low-lying state
\emph{in the same topological sector} as the ground state. This will establish the existence
of an ``unexpected'' low-lying state, which can plausibly be taken as evidence for a gapless
edge mode.

Because we want to establish this stronger claim, the generalized flux insertion argument begins
by inserting not $\pm \Phi_0/2$ flux but $\pm \Phi_0/2e^*$ flux where $e^*$ is the smallest charged
excitation, in units of $e$. By inserting this (larger) amount of flux, we guarantee that the resulting
states, $|\Psi_1\>, |\Psi_2\>$ lie in the same topological sector.\cite{LevinStern,GefenThouless}
To see this, note that $|\Psi_1\>$ can be obtained from $|\Psi_2\>$ by inserting $\Phi_0/e^*$ flux.
This flux insertion process changes the Berry phase associated with braiding a quasiparticle around the cylinder by
\begin{equation}
\Delta \theta = 2\pi q \cdot \frac{1}{e^*}
\label{berry}
\end{equation}
where $q$ is the charge of the quasiparticle (in units of $e$). By construction $\frac{q}{e^*}$ is always an
integer so that $\Delta \theta$ is a multiple of $2\pi$ for every quasiparticle. Hence, $|\Psi_1\>,
|\Psi_2\>$ have the same Berry phases with respect to all quasiparticles -- implying that
they belong to the same topological sector.

Other then this small modification, the argument proceeds as in the integer case. In the first step, we show that
the state $|\Psi_1\>$ has a Kramers degeneracy near the two ends of the
cylinder as long as $\frac{1}{e^*}\chi^T \mathcal{K}^{-1} \tau$ is odd. In the second step, we use this Kramers
degeneracy to construct a protected low-lying excited state $|\Psi_{ex}\>$ at zero flux. Similarly to the integer case, we construct
$|\Psi_{ex}\>$ by starting with $|\Psi_2\>$ and adiabatically inserting $-\Phi_0/2e^*$ flux. Importantly, 
$|\Psi_{ex}\>$ is guaranteed to be in the same topological sector as the ground state $|\Psi_0\>$, since $|\Psi_1\>,|\Psi_2\>$ are 
in the same topological sector. In this way, we see that if $\frac{1}{e^*}\chi^T \mathcal{K}^{-1} \tau$ is odd, then 
the system has a protected low-lying state in the same topological sector as the ground state. This is exactly what 
we wanted to show.

The only piece of the argument which is missing is the proof that $|\Psi_1\>$ has a Kramers degeneracy near
the two ends of the cylinder whenever $\frac{1}{e^*}\chi^T \mathcal{K}^{-1} \tau$ is odd. We now
establish this fact. The first step is to explain more precisely what it means to have a Kramers
degeneracy near the two ends of the cylinder.

\subsection{Local Kramers degeneracy and a local analogue of Kramers theorem} \label{lockramsec}
In this section, we give a precise definition of ``local Kramers degeneracy.''
We also use this definition to state and prove a local analogue of Kramers theorem.

We start by reviewing the usual
(global) notion of Kramers degeneracy. Recall that a quantum many-body
state  $|v\>$ is said to be ``Kramers degenerate'' with its time reversed partner
$|v'\>= \mathcal{T} |v\>$ if $|v\>$ contains an odd number of electrons,
or equivalently $\mathcal{T}^2 |v\> = -|v\>$. The motivation for this
terminology is that, if $v, v'$ are of this form, then it follows that
\begin{align}
\< v'| \mathcal{O} | v\> = 0 \ , \ \<v| \mathcal{O} |v\> = \<v'| \mathcal{O} | v'\>
\label{kramthm}
\end{align}
for any Hermitian time reversal invariant operator $\mathcal{O}$. In particular,
$|v\>, |v'\>$ are guaranteed to be orthogonal and degenerate in energy for any time reversal
invariant Hamiltonian $H$. The result (\ref{kramthm}) is known as Kramers theorem.

Now suppose that $|v\>$ is a quantum many body state with an \emph{even} number of electrons.
In this case, $|v\>$ does not satisfy the requirements for the usual global Kramers
degeneracy. However, $|v\>$ may still exhibit a ``local Kramers degeneracy.'' Imagine,
for example, we take a time reversal invariant insulator and insert two
additional electrons, trapping them in widely separated potential wells.
Let us denote the position of the two potential wells by $a,b$.
We expect intuitively that we can treat the two regions near $a, b$ as two separate systems,
each with an odd number of electrons and a corresponding local Kramers degeneracy.
Moreover, we expect that this local Kramers degeneracy guarantees that the ground state is part
of a multiplet of $4$ degenerate states -- $2$ states coming from each region.

We now formalize this intuition. First, we give a definition of local Kramers degeneracy;
afterwards we state a local analogue of Kramers theorem. All proofs are given in the appendix.

First, we need to define the notion of a ``local operator'': we will say that an operator is local 
if it can be written as a sum of terms, each of which is a product of an
even number of electron creation and annihilation operators acting in some
finite sized region. One implication of this definition is that local operators always 
commute if they act on non-overlapping regions.

We are now ready to define ``local Kramers degeneracy.'' Let $|v\>$ be a quantum many body state 
with an even number of electrons. Suppose $v$ has short range correlations. That is,
\begin{equation}
\<v|\mathcal{O}_1 \mathcal{O}_2 |v\> = \<v|\mathcal{O}_1 |v\> \<v| \mathcal{O}_2 |v\>
\label{lkd0}
\end{equation}
for any two widely separated local operators, $\mathcal{O}_1, \mathcal{O}_2$.
We will say that $|v\>$ has a local Kramers degeneracy in regions $a, b$ if:
\begin{enumerate}
\item{
The state $|v\>$ satisfies
\begin{equation}
\mathcal{T} |v\> = S_a S_b |v\>
\label{lkd1}
\end{equation}
for some local operators $S_a, S_b$ acting on regions $a, b$. We will assume $S_a,S_b$ 
are normalized so that $\|S_a v\| = \|S_b v\| = 1$.
}
\item{
The state $|v\>$ satisfies
\begin{equation}
\mathcal{T}_a^2 |v\> = \mathcal{T}_b^2 |v\> = -|v\>
\label{lkd2}
\end{equation}
where $\mathcal{T}_a \equiv \mathcal{T} S_b$, $\mathcal{T}_b \equiv \mathcal{T} S_a$.
}
\end{enumerate}

We now explain the physical meaning of these conditions. The first condition
(\ref{lkd1}) is essentially the statement that $|v\>$ is time reversal invariant away from regions
$a, b$. The second condition (\ref{lkd2}) formalizes the notion of having an odd number of electrons 
localized in regions $a, b$. The basic idea is that the two anti-linear operators $\mathcal{T}_a , \mathcal{T}_b$
implement a local time reversal symmetry transformation in the two regions $a,b$.
The condition $\mathcal{T}_a^2 |v\> = -|v\>$ is then analogous to the usual requirement
for Kramers theorem, $\mathcal{T}^2 |v\> = -|v\>$.

A useful fact, proved in appendix \ref{twopossapp}, is that if $|v\>$ satisfies
(\ref{lkd0}), (\ref{lkd1}), then either
\begin{align}
\mathcal{T}_a^2 |v\> = \mathcal{T}_b^2 |v\> = +|v\> \ \text{ or } \ \mathcal{T}_a^2 |v\> = \mathcal{T}_b^2 |v\> = -|v\>
\label{twopossrel}
\end{align}
The two cases correspond (roughly speaking) to having either an even or odd number
of electrons localized near regions $a, b$. (Not surprisingly, local Kramers
degeneracies always come in pairs, since we have assumed that the total number of electrons is even).

To get a feeling for this definition, it is useful to think about the above example of a time
reversal invariant insulator state $|v\>$, with two additional electrons localized near points $a, b$.
Intuitively, $|v\>$ has a local Kramers degeneracy near $a, b$. At the same time, $|v\>$ satisfies the above
conditions, as we now show. To establish the first condition (\ref{lkd1}), let us assume without loss of generality 
that the two electron spins point in the $+\hat{z}$ direction. Then, we have $\mathcal{T} |v\> = \sigma^x_a \sigma^x_b |v\>$ where
$\sigma^x_a, \sigma^x_b$ are spin-flip operators acting on the spins localized near $a,b$. Hence, we can
take $S_a = \sigma^x_a$, $S_b = \sigma^x_b$. As for the second condition (\ref{lkd2}), this relation follows from
the fact that $\mathcal{T} \sigma^x_b \mathcal{T} = -\sigma^x_b$ so that
$\mathcal{T}_a^2 |v\> = (\mathcal{T} \sigma^x_b)^2 |v\> = -|v\>$ (and similarly for $\mathcal{T}_b^2 |v\>$). Finally, we note that
$\mathcal{T}_a$ does in fact implement a local time reversal transformation near region $a$:
$\mathcal{T}_a |v\>$ is a state where the spin near $b$ points in the
$+\hat{z}$ direction and the spin near $a$ points in the $-\hat{z}$ direction -- the ``local time
reverse'' of $|v\>$.

To complete our discussion, we now state a local analogue of Kramers theorem (\ref{kramthm}) based on this
definition. The proof of this result is given in appendix \ref{lockrapp}. 

{\bf Local Kramers theorem:} \emph{Let $|v\>$ be a quantum many body
state satisfying conditions (\ref{lkd0}-\ref{lkd2}) so that $|v\>$ has a local Kramers degeneracy in regions $a, b$.
Let $|v'\> = \mathcal{T}_a |v\>$. Then $|v\>, |v'\>$ 
satisfy
\begin{align}
\<v'|\mathcal{O}|v\>  = 0 \ , \ \<v|\mathcal{O}|v\> = \<v'|\mathcal{O}|v'\>
\label{lockramthm}
\end{align}
for any $\mathcal{O}$ which is a finite product of local, Hermitian, time reversal invariant operators.}

In order to understand the implications of this result, suppose that $|v\>$ is a ground state of a time reversal invariant Hamiltonian $H$
with a finite energy gap. We can see from (\ref{lockramthm}) that $|v'\> = \mathcal{T}_a |v\>$ is orthogonal to $|v\>$ and degenerate 
in energy. In other words, (\ref{lockramthm}) guarantees that $H$ has a two-fold ground state degeneracy. In addition, (\ref{lockramthm}) implies that if we add an arbitrary local time reversal invariant perturbation to $H$ then the degeneracy between $|v\>,|v'\>$ does not 
split to any finite order in perturbation theory: at each order, the off-diagonal matrix elements vanish, while the two diagonal elements 
are identical. In this way, (\ref{lockramthm}) implies the existence of a robust ground state degeneracy which is analogous to the usual Kramers degeneracy.

It is worth mentioning that while the above theorem focuses entirely on the Kramers degeneracy in region $a$, 
there is an identical two-fold degeneracy
coming from region $b$. Thus, there are four degenerate states altogether: 
$|v\>,|v'\> = \mathcal{T}_a |v\>, |v''\> = \mathcal{T}_b |v\>, |v'''\> = \mathcal{T} |v\>$.
This entire multiplet of four states obeys the analogue of Eq. (\ref{lockramthm}), as can be shown using arguments similar to the 
ones given in appendix \ref{lockrapp}.

\subsection{Establishing that \texorpdfstring{$|\Psi_1\>$}{|Psi1>} has a local Kramers degeneracy}
We now fill in the missing piece of the flux insertion argument from section 
\ref{genfluxargsec}: we show that if $\frac{1}{e^*} \chi^T \mathcal{K}^{-1} \tau$ is odd,
then $|\Psi_1\>$ has a local Kramers degeneracy near the two ends of the cylinder.

The first step is to understand the relationship between the two states $|\Psi_1\>, |\Psi_2\>$. Recall that
these states are obtained by starting in the state $|\Psi_0\>$ and adiabatically inserting $\pm \Phi_0/2e^*$ flux. 
Therefore, if we start in $|\Psi_1\>$ and then 
insert -$\Phi_0/e^*$ flux, we obtain $|\Psi_2\>$.

We next use the edge theory (\ref{kedge}) for the two edges of the cylinder to analyze the effect
of this flux insertion process. A simple calculation (see appendix \ref{fluxapp}) shows that
the effect of the flux insertion process is given by applying an operator $S_l \cdot S_r$ to $|\Psi_1\>$, where
$S_l$ acts on the left edge and $S_r$ acts on the right edge:
\begin{equation}
|\Psi_2\> = S_l S_r |\Psi_1\>
\label{fluxeffect}
\end{equation}
Here $S_l, S_r$ are given by
\begin{align}
S_l = \Gamma_l(\Lambda_c) \ , \ S_r = \Gamma_r(-\Lambda_c)
\end{align}
with 
\begin{equation}
\Gamma_l(\Lambda) \equiv \int \frac{dx}{\sqrt{L}} e^{i\Theta_l(\Lambda)}
\end{equation}
and similarly for $\Gamma_r$. The vector $\Lambda_c$ is defined by $\Lambda_c = \frac{1}{e^*} \mathcal{K}^{-1} \tau$.

We can now identify $S_l, S_r$ with the $S_a, S_b$ operators
from the definition of local Kramers degeneracy. With this identification, equation (\ref{fluxeffect})
immediately establishes one of the conditions (\ref{lkd1}) for local Kramers degeneracy. All that remains
is to prove the relation (\ref{lkd2}). In other words, we need to show
$\mathcal{T} S_l \mathcal{T} S_l |\Psi_1\> = - |\Psi_1\>$. To this end, note that
according to (\ref{timereversal}), $\Gamma_l(\Lambda_c)$ transforms under time reversal as
\begin{equation}
\mathcal{T}^{-1} \Gamma_l(\Lambda_c) \mathcal{T} = \Gamma_l(-\Lambda_c) e^{i \pi \chi^T \Lambda_c}
\end{equation}
Then, since $\chi^T \Lambda_c = \frac{1}{e^*} \chi^T \mathcal{K}^{-1} \tau$ is odd, we have
\begin{equation}
\mathcal{T}^{-1} \Gamma_l(\Lambda_c) \mathcal{T} = -\Gamma_l(-\Lambda_c)
\end{equation}
It follows that
\begin{eqnarray}
\mathcal{T} S_l \mathcal{T} S_l |\Psi_1\>
&=& - \Gamma_l(-\Lambda_c) \Gamma_l (\Lambda_c) |\Psi_1\> \nonumber \\
&=& -\Psi_1
\end{eqnarray}
as required.

\section{Conclusion}
In this work, we have investigated the properties of general time reversal invariant
insulators with abelian quasiparticle statistics. First, we constructed all possible
Chern-Simons theories that can describe such insulators. Second, we derived a general
criterion for when such states have protected edge modes -- that is, edge modes that
cannot be gapped without breaking time reversal or charge conservation symmetry, explicitly
or spontaneously. Finally, we gave a precise
definition of ``local Kramers degeneracy'' and we proved a local analogue of Kramers theorem
-- important concepts in the theory of topological insulators.

A number of questions remain open. First, we do not yet have microscopic realizations
of all the Chern-Simons theories discussed here. In particular, we do not have any examples
where the vector $x$ in the definition of $\chi$ (\ref{gensolT}) is nonzero. Such examples
would be particularly interesting because, in their simplest form, they would give microscopic
models for bosonic topological insulators without fractionalization (see section \ref{examplesec}
for a brief discussion). These bosonic topological insulators were conjectured to exist in 
Ref. [\onlinecite{XieSPT4}], but have not yet been studied in the context of concrete 
microscopic models.

Another issue is that we have focused entirely on insulators with \emph{abelian} statistics.
Yet it is not hard to construct microscopic models for non-abelian time reversal invariant insulators.
For example, one can imagine toy models, similar to those discussed in Ref. [\onlinecite{LevinStern}],
where the spin-up and spin-down electrons each form \emph{non-abelian} fractional quantum Hall states
with opposite chiralities. It would be interesting to analyze the stability of the edge modes in this case. This stability
question was partially addressed by the flux insertion argument in Ref. [\onlinecite{LevinStern}]. However, the analysis
in Ref. [\onlinecite{LevinStern}] is not yet complete since the flux insertion argument only allows us to prove that 
the edge modes are protected when $\sigma_{sH}/e^*$ is odd; it does not prove that the edge modes can be gapped out when $\sigma_{sH}/e^*$
is even. Completing this analysis requires a microscopic investigation of the stability of the edge modes,
and is an interesting question for future research.

Finally, it would be interesting to generalize the approach presented here to the three dimensional case. We now have a number of
examples\cite{LevinBurnellKoch,Swingle,Maciejko} of three dimensional fractional topological insulators,
but these constructions almost certainly do not exhaust all the possibilities. It would be interesting to develop
a general classification scheme for 3D fractional topological insulators analogous to the one discussed here.

\acknowledgments
ML was supported in part by an Alfred P. Sloan Research Fellowship. AS thanks the US-Israel Binational Science
Foundation, the Minerva foundation and Microsoft Station Q for financial support.

\appendix

\section{General time reversal invariant Chern-Simons theories} \label{genabelapp}
In this section we find the most general $p \times p$ integer matrices, $\mathcal{K}, T$ and $p$ component vectors
$\tau, \chi$ satisfying
\begin{eqnarray}
T^T \mathcal{K} T &=& - \mathcal{K}  \label{ktcondapp} \\
T^2 &=& 1	\label{t2condapp} \\
T \tau &=& \tau \label{tchgcondapp} \\
(1-T^T) \chi &\equiv& \tau \text{ (mod $2$)} \label{tchicondapp}
\end{eqnarray}
We show that the most general solutions to these consistency conditions are of the form (\ref{gensolK}), (\ref{gensolT}),
up to the equivalence transformations (\ref{uequiv}), (\ref{chiequiv}), (\ref{chiequiv2}).

To begin, we note that $Tr(T) =0$. We can derive this fact by
multiplying both sides of (\ref{ktcondapp}) on the right by $T \mathcal{K}^{-1}$ and taking the trace.

Next, we note that $T$ has eigenvalues $\pm 1$ since $T^2 = 1$. Combining this with
the fact that $Tr(T) =0$, we conclude that $T$ has an equal number of $+1$
and $-1$ eigenvalues. Let the number of $+1$ eigenvalues be $N$, 
so that $p = 2N$. Interestingly, we can already see that $p$ must be even. 

In the third step, we choose a basis $\{v_1,...,v_N\}$ for the $+1$ eigenspace of $T$. We choose this basis so that the
$v_i$ are integer vectors. This is always possible since this eigenspace is spanned by the
columns of $1+T$, a matrix with integer entries. In fact, we will go a step further and choose the basis so that
the $N \times N$ minors of the matrix with columns $\{v_1,...,v_N\}$ have no common factor. This property guarantees
that we can extend $\{v_1,...,v_N\}$ to an integer basis $\{v_1,...,v_N,w_1,...,w_N\}$ for the whole $p = 2N$ dimensional space,
such that the matrix with columns $\{v_1,...,v_N,w_1,...,w_N\}$ has determinant $\pm 1$.

We next make a change of basis to $\{v_1,...,v_N,w_1,...,w_N\}$. Equivalently, we make a transformation $T \rightarrow U T U^{-1}$
(\ref{uequiv}) where $U^{-1}$ is the matrix with columns $\{v_1,...,v_N,w_1,...,w_N\}$. After this change of basis, $T$ is of the form
\begin{equation}
T = \bpm \bf{1}_N & F \\
	   0 & G \epm
\end{equation}
where $F,G$ are of dimensions $N \times N$.

In the fifth step, we use $T^2 = 1$ to deduce that $G^2 = 1$.  We also have $Tr(G) = Tr(T)-N = -N$.
Combining these two facts, we conclude that all the eigenvalues of $G$ are equal to $-1$
so that $G$ = $-\bf{1}_{N}$. Hence, $T$ is of the form
\begin{equation}
T = \bpm \bf{1}_N & F \\
	   0 & -\bf{1}_N \epm
\end{equation}

The sixth step is to make another transformation $T \rightarrow U T U^{-1}$ (\ref{uequiv}), where $U$ is
an integer matrix of the form
\begin{equation}
U = \bpm U_1 & 0 \\
	 0 &  U_2 \epm
\end{equation}
and $\det(U_1) = \det(U_2) = \pm 1$. Under this transformation, $F \rightarrow U_1 F U_2^{-1}$. By choosing $U_1, U_2$
appropriately we can make $F$ a diagonal matrix: this follows from the Smith normal form\cite{Prasolov} for integer matrices.

The seventh step is to make another transformation $T \rightarrow U T U^{-1}$ (\ref{uequiv}), where $U$ is of
the form
\begin{equation}
U = \bpm \bf{1}_N & Y \\
	  0 &	\bf{1}_{N} \epm
\end{equation}
Under this transformation, $F \rightarrow F - 2Y$. By choosing $Y$ appropriately, we can guarantee that $F$ has only
$0$'s and $1$'s on the diagonal. Hence, we can assume without loss of generality that $F$ is of the form
\begin{equation}
F = \bpm \bf{1}_{N-M} & 0 \\
	  0	& 0 \epm
\end{equation}
where $M \leq N$. Putting this all together, we conclude that $T$ can be written in the form
\begin{equation}
T = \bpm \bf{1}_{N-M} & 0 & \bf{1}_{N-M} & 0 \\
	  0 & \bf{1}_{M} & 0 & 0 \\
	  0 &  0 &	-\bf{1}_{N-M} & 0 \\
	  0 & 0 &	0	& -\bf{1}_{M} \epm
\end{equation}

The final step is to make yet another transformation $T \rightarrow U T U^{-1}$ (\ref{uequiv}), where $U$ is of the
form
\begin{equation}
U = \bpm \bf{1}_{N-M} & 0 & 0 & 0 \\
	  0 & \bf{1}_{M} & 0 & 0 \\
	  \bf{1}_{N-M} & 0 & \bf{1}_{N-M} & 0 \\
	   0 	& 0	& 0 & \bf{1}_{M} \epm
\end{equation}
This transformation changes $T$ to
\begin{equation}
T = \bpm 0 & 0 & \bf{1}_{N-M} & 0 \\
	0  & \bf{1}_{M} & 0 & 0 \\
	\bf{1}_{N-M} & 0 & 0 & 0 \\
	0 & 0 & 0 & -\bf{1}_{M} \epm
\end{equation}
After reordering the rows and columns, we arrive at
\begin{equation}
T = \bpm -\bf{1}_{M} & 0 & 0 & 0 \\
	     0 & \bf{1}_{M} & 0 & 0 \\
	     0 & 0 & 0 & \bf{1}_{N-M} \\
	     0 & 0 & \bf{1}_{N-M} & 0 \epm
\end{equation}

Using the relations (\ref{ktcondapp}), (\ref{tchgcondapp}) we then deduce that $\mathcal{K}, \tau$ must be of the form
\begin{align}
\mathcal{K} = \bpm 0 & A & B & B \\
		   A^T & 0 & C & -C \\
		   B^T & C^T & K & W \\
		   B^T & -C^T & W^T & -K \epm \ , \ \tau = \bpm 0 \\ t' \\ t \\ t \epm
\end{align}
where matrix $A$ is of dimension $M \times M$, $\mathcal{K}$ and $W = -W^T$ are of dimension $(N-M) \times (N-M)$ and the matrices
$B,C$ are of dimension $M \times (N-M)$. Also, $t'$ is an $M$ component integer vector and $t$ is an $(N-M)$
component integer vector. In fact, it follows from condition (\ref{paritycond}) that $t'$ is an \emph{even} vector.

As for $\chi$, it is not hard to see that we can put $\chi$ in the form
\begin{equation}
\chi = \bpm y_1 \\ 0 \\ 0 \\ y_2 \epm
\end{equation}
using an appropriate transformation $\chi \rightarrow  \chi + \frac{1}{\pi}\mathcal{K}(1-T)\xi$ (\ref{chiequiv}).
Then, using the relation (\ref{tchicondapp}), we deduce that
\begin{equation}
y_2 \equiv t \text{ (mod $2$)}
\end{equation}
and $y_1$ is an integer vector.
Finally, we use the equivalence $\chi \rightarrow \chi + 2v$ (\ref{chiequiv2}) to set $y_2 = t$, and to change $y_1$
into a vector of $1$'s and $0$'s. This completes our derivation of (\ref{gensolK}), (\ref{gensolT}).

\section{Reformulation of edge mode criterion in terms of spin-Hall conductivity} \label{shcritapp}
According to the results in sections \ref{microedgesec}-\ref{fluxargsec},
the insulators described by (\ref{gensolK}), (\ref{gensolT}) have
protected edge modes if and only if $\frac{1}{e^*} \chi^T \mathcal{K}^{-1} \tau$
is odd. In this section, we show that this criterion can be reformulated in simpler
language for a large class of $s^z$ conserving insulators.
To be specific, we consider insulators such that $\chi \equiv \tau_\down \text{ (mod $2$)}$ where
$\tau_\up,\tau_\down$ are the ``charge vectors'' which keep track of the (separately conserved) spin-up, spin-down electrons.
For this class of insulators, we show that the edge mode criterion can be equivalently phrased in terms of the parity of
$\sigma_{sH}/e^*$ where $\sigma_{sH}$ is the spin-Hall conductivity in units of $e/2\pi$.

To see this, define
\begin{equation}
\Lambda_c = \frac{1}{e^*} \mathcal{K}^{-1} \tau
\end{equation}
Then, we can write
\begin{equation}
\frac{1}{e^*} \chi^T \mathcal{K}^{-1} \tau = \chi^T \Lambda_c
\label{chicrit}
\end{equation}
At the same time, we have
\begin{equation}
\sigma_{sH} = \frac{1}{2} (\tau_\up - \tau_\down)^T\mathcal{K}^{-1}\tau   = -\tau_\down^T\mathcal{K}^{-1} \tau
\end{equation}
where the second equality follows from time reversal symmetry. Hence,
\begin{equation}
\frac{\sigma_{sH}}{e^*} = -\frac{1}{e^*} \tau_\down^T \mathcal{K}^{-1} \tau = -\tau_\down^T \Lambda_c
\label{shcrit}
\end{equation}
Comparing (\ref{chicrit}), (\ref{shcrit}), the claim follows immediately: if $\chi \equiv \tau_\down \text{ (mod $2$)}$,
then $\frac{1}{e^*} \chi^T \mathcal{K}^{-1} \tau$ has the same parity as $\frac{\sigma_{sH}}{e^*}$.
Thus, in this case, the criterion for protected edge modes can be equivalently formulated as the condition
that $\frac{\sigma_{sH}}{e^*}$ is odd.

\section{The null vector criterion} \label{nullvecapp}
In this section, we explain why the null vector condition (\ref{cond1}) guarantees that the
perturbation (\ref{pert4}) can gap out the two pairs
of edge modes in Example 2 (section \ref{twocompsec}). The idea is as follows: we consider
a change of variables of the form $\Phi' = U \Phi$
where $U$ is a $4 \times 4$ matrix whose first two rows are $\Lambda_1^T \mathcal{K}, \Lambda_2^T \mathcal{K}$.
The condition (\ref{cond1}) guarantees that, if we choose the other two rows of $U$ appropriately,
then the resulting $\mathcal{K}' = (U^{-1})^T\mathcal{K} U^{-1}, \tau' = (U^{-1})^T \tau$ can be put in the form
\begin{equation}
\mathcal{K}' = \bpm 0 & 0 & 1 & 0 \\
		    0 & 0 & 0 & 1 \\
		    1 & 0 & 0 & 0 \\
		    0 & 1 & 0 & 0 \epm , \ \tau' = \bpm t_1 \\ t_2 \\ 0 \\ 0 \epm
\end{equation}
(The reason that $\tau'$ takes the above form is that $\Lambda_1, \Lambda_2$ are \emph{charge conserving} vectors).
 
We next assume that interactions on the edge are such that $\mathcal{V'} = (U^{-1})^T\mathcal{V} U^{-1}) = v \delta_{IJ}$
(we can make this assumption without loss of generality since the velocity matrix is non-universal and can
be modified by appropriate perturbations at the edge). In this case, the edge theory can be written as the sum
of two decoupled actions:
\begin{equation}
L = L_{13} + L_{24}
\end{equation}
where
\begin{eqnarray}
L_{13} &=& \frac{1}{4\pi} \left(2\partial_x \Phi'_1 \partial_t \Phi'_3
- v (\partial_x\Phi'_1)^2 - v (\partial_x \Phi'_3)^2 \right) \nonumber \\
&+& \frac{t_1}{2\pi} \epsilon^{\mu \nu} \partial_\mu \Phi'_1 A_\nu
\end{eqnarray}
and
\begin{eqnarray}
L_{24} &=& \frac{1}{4\pi} \left(2\partial_x \Phi'_2 \partial_t \Phi'_4
- v (\partial_x\Phi'_2)^2 - v (\partial_x \Phi'_4)^2 \right) \nonumber \\
&+& \frac{t_2}{2\pi} \epsilon^{\mu \nu} \partial_\mu \Phi'_2 A_\nu
\end{eqnarray}

In the new $\Phi'$ variables, the perturbation (\ref{pert4}) becomes:
\begin{equation}
U(x)[\cos(\Phi_1'-\alpha(x)) + (-1)^{Q(\Lambda_1)}\cos(\Phi_2'-\alpha(x))]
\label{pertnew}
\end{equation}
It is now easy to analyze the effect of the perturbation (\ref{pertnew}): clearly, this
term will gap out the two Luttinger liquids by freezing the values of $\Phi_1', \Phi_2'$.
Note that the relevance or irrelevance of (\ref{pertnew}) is not important here, since the 
system will always be gapped out for large $U$. Alternatively, it is not hard to see that 
we can always make (\ref{pertnew}) relevant if we tune the velocity matrix
$\mathcal{V'}$ so that the coefficients of $(\partial_x \Phi'_1)^2$, $(\partial_x \Phi'_2)^2$
are much larger than the coefficients of $(\partial_x \Phi'_3)^2$, $(\partial_x \Phi'_4)^2$ .

\section{Primitivity condition} \label{primcondapp}
Let $\Lambda_1,...,\Lambda_N$ be $N$ integer vectors with $M$ components, $M \geq N$. 
Let $B$ denote the matrix with columns $\Lambda_1,...,\Lambda_N$. In this section, we 
show that there exist integers $a_1,...,a_N$ with no common divisor such that the
linear combination $\sum_i a_i \Lambda_i$ is non-primitive, if and only if
the set of $\bpm 2N \\ N \epm$ $N \times N$ minors of the matrix $B$ have a common factor. 

To begin, we make use of the Smith normal form\cite{Prasolov} for integer matrices. According
to this result, we can always find an $M \times M$ integer matrix $S$ and an 
$N \times N$ integer matrix $T$, both with determinant $\pm 1$, such that $B = S \cdot D \cdot T$ 
where $D$ is an $M \times N$ integer matrix of the form
\begin{equation}
D = \bpm d_1 & 0 & \cdots & 0 \\
	 0  & d_2 & \cdots & 0 \\
	 \vdots & \vdots & \vdots & \vdots \\
	 0  & 0 & \cdots & d_N \\ 
         \vdots & \vdots & \vdots & \vdots \\
	0 & 0 & \cdots & 0 \epm	
\label{D}
\end{equation}
Next, we compare the $N \times N$ minors of $B$ and $D$. Let $b$ denote the greatest common factor of the 
$N \times N$ minors of $B$, and let $d$ denote the greatest common factor of the $N \times N$ minors of $D$. Given that 
$S$ and $T$ are integer matrices, it follows from $B = S \cdot D \cdot T$ that the minors of $B$ are integer linear 
combinations of the minors of $D$. Hence, $d$ divides into $b$. On the other hand, since $S$ and $T$ have determinant 
$\pm 1$, we can equally well write $D = S^{-1} \cdot B \cdot T^{-1}$, implying that the minors of $D$ are integer 
linear combinations of the minors of $B$. Thus, we also know that $b$ divides into $d$. Combining these two observations gives $b=d$. 
At the same time, the explicit expression for $D$ (\ref{D}) immediately implies $d= |d_1 \cdot d_2 \cdots d_N |$. We conclude that the
greatest common factor $b$ is larger than $1$ if and only if one of the $d_j$'s is different from $\pm 1$. 

To complete the argument, we need to show that having one of the $d_j$'s different from $\pm 1$ is necessary and 
sufficient for having a non-primitive linear combination $\sum_i a_i \Lambda_i$. 
To see that this condition is sufficient, suppose that one of the $d_j$'s is different from $\pm 1$. We then define
$a_1,...,a_N$ to be the elements of the $j$th column of $T^{-1}$. Using $B = S \cdot D \cdot T$, we can see that 
$\sum_i a_i \Lambda_i = d_j \cdot \Lambda$, where $\Lambda$ is the $j$th column of $S$. Also, we can see that the $a_i$ have no 
common divisor since $T^{-1}$ has determinant $\pm 1$. Hence, we have constructed a non-primitive linear combination. Conversely, suppose 
$d_i = \pm 1$ for all $i$, and let $a_1,...,a_N$ be integers with no common factor. Defining $a = (a_1,...,a_N)$, we have
$\sum_i a_i \Lambda_i = B \cdot a = S \cdot D \cdot T \cdot a$. We then note that since $S, T$ have determinant
$\pm 1$, the vector $S \cdot D \cdot T \cdot a$ must have no common factor. We conclude that $\sum_i a_i \Lambda_i$ is
primitive for any choice of $a_i$. This completes the proof.

\section{Proof of Eq. \ref{twopossrel} regarding local Kramers degeneracies} \label{twopossapp}
Let $|v\>$ be a quantum many body state with an even number of electrons and short range correlations.
In this section, we show that if $|v\>$ satisfies (\ref{lkd1}), then either
\begin{equation}
\mathcal{T}_a^2 |v\> = \mathcal{T}_b^2 |v\> = |v\> \ \text{ or } \ \mathcal{T}^2_a |v\> = \mathcal{T}^2_b |v\> = -|v\>
\end{equation}
The first step is to show that $\mathcal{T}_a^2 |v\> = \zeta |v\>, \mathcal{T}^2_b |v\> = \frac{1}{\zeta} |v\>$
for some complex number $\zeta$. To see this, we define $A = \mathcal{T}_a^2$, $B = \mathcal{T}_b^2$,
and note that
\begin{eqnarray}
BA |v\> &=& (\mathcal{T} S_a \mathcal{T} S_a)(\mathcal{T} S_b \mathcal{T} S_b) |v\> \nonumber \\
&=& (\mathcal{T} S_a \mathcal{T}) S_a (\mathcal{T} S_b \mathcal{T}) S_b |v\> \nonumber \\
&=& (\mathcal{T} S_a \mathcal{T})(\mathcal{T} S_b \mathcal{T}) S_a S_b |v\> \nonumber \\
&=& \mathcal{T} S_a S_b \mathcal{T} S_a S_b |v\> \nonumber \\
&=& |v\>
\label{BAid}
\end{eqnarray}
Here the third equality follows from the fact that $S_a, (\mathcal{T} S_b \mathcal{T})$ are
local operators acting in non-overlapping regions and therefore commute. Next we note:
\begin{eqnarray}
\<v|B^\dagger B |v\>\<v| A^\dagger A |v\> &=& \<v| (B^\dagger B) (A^\dagger A) |v \> \nonumber \\
&=& \<v| A^\dagger B^\dagger B A |v\> \nonumber \\
&=& \<v | v\>  \nonumber \\
&=& 1
\label{eqab}
\end{eqnarray}
where the first equality comes from the fact that $v$ has short range
correlations and $A,B$ are local operators acting in distant locations:
$A = \mathcal{T}_a^2 = (\mathcal{T} S_b \mathcal{T}) S_b$ acts in region
$b$, while $B = \mathcal{T}_b^2 = (\mathcal{T} S_a \mathcal{T}) S_a$
acts in region $a$.

On the other hand, if we insert a complete set of states
$\Sigma_i |v_i\>\<v_i|$ into $\<v|B^\dagger B |v\>,\<v|A^\dagger A |v\>$ and we choose this
set of states so that one of the $v_i$'s is $v$, we deduce the
inequalities
\begin{eqnarray}
\<v|B^\dagger B |v\> &\geq& |\<v|B|v\>|^2 \nonumber \\ 
\<v|A^\dagger A |v\> &\geq& |\<v|A|v\>|^2
\end{eqnarray}
Multiplying these two inequalities gives
\begin{eqnarray}
\<v|B^\dagger B |v\>\<v|A^\dagger A|v\> &\geq& |\<v|B|v\>|^2 \cdot |\<v|A|v\>|^2 \nonumber \\
&=& |\<v|BA|v\>|^2 \nonumber \\
&=& |\<v | v\>|^2 \nonumber \\
&=& 1
\label{ineqab}
\end{eqnarray}
where the first equality again follows from the fact that $v$ has short
range correlations. Comparing (\ref{eqab}) and (\ref{ineqab}), we deduce that $v$ must be an eigenvector of both
$A,B$, since this is the only way the inequality can be saturated. In other
words, $\mathcal{T}_a^2 |v\> = \zeta |v\>, \mathcal{T}_b^2 |v\> = \frac{1}{\zeta} |v\>$ for some
complex $\zeta$.

Now we show that $\zeta = \pm 1$. First, we note that
\begin{eqnarray}
\mathcal{T}_a |v\> &=& \mathcal{T} S_b |v\> \nonumber \\
&=& \mathcal{T} S_b (\mathcal{T} S_a S_b |v\>) \nonumber \\
&=& (\mathcal{T} S_b \mathcal{T}) S_a S_b |v\> \nonumber \\
&=& S_a (\mathcal{T} S_b \mathcal{T}) S_b |v\> \nonumber \\
&=& S_a \mathcal{T}_a^2 |v\> 
\label{saidentity}
\end{eqnarray}
so that
\begin{equation}
\mathcal{T}_a |v\> = \zeta S_a |v\>
\label{saidentity2}
\end{equation}
This relation, together with the fact that $\|\mathcal{T}_a v\| =
\|\mathcal{T} S_b v\| = \|S_b v\| = 1$, and $\|S_a v\| = 1$, implies
that $|\zeta| = 1$. To see that $\zeta = \pm 1$, we note that
\begin{equation}
\mathcal{T}_a^2 (\mathcal{T}_a |v\>) = \mathcal{T}_a (\mathcal{T}_a^2 |v\>) = \zeta^* \mathcal{T}_a |v\>
\end{equation}
by anti-linearity. At the same time,
\begin{equation}
\mathcal{T}_a^2 (S_a |v\>) = S_a \mathcal{T}_a^2 |v\> = \zeta S_a |v\>
\end{equation}
where the first equality follows from the fact that $\mathcal{T}_a^2, S_a$
act in different regions and hence commute. Given that $\mathcal{T}_a |v\> \propto S_a |v\>$
by (\ref{saidentity2}), we must have $\zeta = \zeta^*$. We conclude that $\zeta = \pm 1$.

\section{Proof of the local analogue of Kramers theorem} \label{lockrapp}
In this section we prove the local analogue of Kramers theorem described
in section \ref{lockramsec}. Let $|v\>$ be a quantum many body state with an even number of electrons, short range correlations, 
and with a local Kramers degeneracy in regions $a,b$  (i.e., suppose $|v\>$ satisfies conditions (\ref{lkd0}-\ref{lkd2})).
Define $|v'\> = \mathcal{T}_a |v\>$ where $\mathcal{T}_a = \mathcal{T} S_b$. The result we wish to establish is that
\begin{eqnarray}
\<v'|\mathcal{O}|v\> &=& 0 \label{lockd1} \\
\<v| \mathcal{O} |v\> &=& \<v'| \mathcal{O} |v'\> \label{lockd2}
\end{eqnarray}
for any $\mathcal{O}$ which is a finite product of local, Hermitian, time reversal invariant operators. 

To begin, we write $\mathcal{O} = \mathcal{O}_1 \cdots \mathcal{O}_k$ where the $\mathcal{O}_i$ are local, 
Hermitian, time reversal invariant operators acting on widely separated regions. We first consider
the special case where all of the $\mathcal{O}_i$ act on regions that are far from $b$. (Here, when we
say ``far'', we mean much farther than the correlation length). The proof in this case is based on 
three observations:
\begin{enumerate}
\item{
The operator $\mathcal{O}$ commutes with $\mathcal{T}_a$.
}
\item{
For any two operators $\mathcal{O}, \mathcal{O'}$ that act on regions far from $b$, 
\begin{equation}
\<\mathcal{T}_a \mathcal{O}' v | \mathcal{T}_a \mathcal{O} v \> = \<\mathcal{O} v | \mathcal{O}' v \>
\label{antiunit}
\end{equation}
In other words, $\mathcal{T}_a$ behaves like an anti-unitary operator within the subspace of states
of the form $\{\mathcal{O} |v\>\}$.
}
\item{The state $|v'\>$ can be written as
\begin{equation}
|v'\> = \mathcal{T}_a |v\> = - S_a |v\>
\label{saidentity3}
\end{equation} 
}
\end{enumerate}
The first observation follows from the fact that $\mathcal{O}$ commutes with $S_b$, and therefore also 
commutes with $\mathcal{T}_a = \mathcal{T} S_b$. The second observation (\ref{antiunit}) follows from
\begin{eqnarray}
\<\mathcal{T}_a \mathcal{O}' v | \mathcal{T}_a \mathcal{O} v\> &=&
\<\mathcal{T} S_b \mathcal{O}' v | \mathcal{T} S_b \mathcal{O} v\> \nonumber \\
&=& \<S_b \mathcal{O} v | S_b \mathcal{O}' v\> \nonumber \\
&=& \<v | \mathcal{O}^\dagger S_b^\dagger S_b \mathcal{O}' v\> \nonumber \\
&=& \<v |(\mathcal{O}^\dagger \mathcal{O}') (S_b^\dagger S_b) | v\> \nonumber \\
&=& \<v |\mathcal{O}^\dagger \mathcal{O}' |v\> \cdot \<v|S_b^\dagger S_b | v\> \nonumber \\
&=& \<v |\mathcal{O}^\dagger \mathcal{O}' |v\> \nonumber \\
&=& \<\mathcal{O} v | \mathcal{O}' v\>
\label{normpres}
\end{eqnarray}
where we are using the fact that $|v\>$ has short range correlations in the fifth equality.
As for the third observation (\ref{saidentity3}), this is equivalent to the identity derived in Eq. \ref{saidentity}.
 
Given these three observations, our argument proceeds just like the proof of the usual Kramers
theorem. To derive (\ref{lockd1}), we note that
\begin{eqnarray}
\<v' |\mathcal{O} v\> &=& \<\mathcal{T}_a \mathcal{O} v | \mathcal{T}_a v'\> \nonumber \\
&=& \<\mathcal{O} \mathcal{T}_a v | \mathcal{T}_a v'\> \nonumber \\
&=& \<\mathcal{O} v' | \mathcal{T}^2_a v\> \nonumber \\
&=& -\<\mathcal{O} v' | v\> \nonumber \\
&=& -\< v' | \mathcal{O} v\> 
\label{case1}
\end{eqnarray}
where the first equality follows from (\ref{antiunit}) and (\ref{saidentity3}). It then follows that
$\<v'| \mathcal{O} v\> = 0$. As for (\ref{lockd2}), we have
\begin{eqnarray}
\<v |\mathcal{O} v\> &=& \<\mathcal{T}_a \mathcal{O} v | \mathcal{T}_a v \> \nonumber \\
&=& \<\mathcal{O} \mathcal{T}_a  v | \mathcal{T}_a v \> \nonumber \\
&=& \<\mathcal{O} v' | v'\> \nonumber \\
&=& \<v' | \mathcal{O} v'\>
\end{eqnarray}

This completes the argument for the special case where all the $\mathcal{O}_i$ act far from $b$. 
We now consider the general case, where some of the $\mathcal{O}_i$ operators may act in the vicinity of $b$. 
We define $\mathcal{O}_\alpha$ to be the product of all the $\mathcal{O}_i$ that act far from $b$, and
$\mathcal{O}_\beta$ to be the product of all the other $\mathcal{O}_i$. We can then express $\mathcal{O}$ as a product
$\mathcal{O} = \mathcal{O}_\alpha \mathcal{O}_\beta$. We note that the definition of $\mathcal{O}_\alpha, \mathcal{O}_\beta$ 
guarantees that $\mathcal{O}_\alpha$ acts on regions far from $b$, while $\mathcal{O}_\beta$ acts on regions far
from $a$. Also, $\mathcal{O}_\alpha ,\mathcal{O}_\beta$ act on regions that are far from one another.

The factorization $\mathcal{O} =\mathcal{O}_\alpha \mathcal{O}_\beta$ is useful because it allows us to 
reduce the general problem to the special case treated above. Indeed, consider the relation (\ref{lockd1}). 
Simple manipulations give 
\begin{eqnarray}
\<v'|\mathcal{O} v\> &=& \<v'|\mathcal{O}_\alpha \mathcal{O}_\beta v\> \nonumber \\
&=& -\<S_a v|\mathcal{O}_\alpha \mathcal{O}_\beta v\> \nonumber \nonumber \\
&=& -\<v|S_a^\dagger \mathcal{O}_\alpha \mathcal{O}_\beta | v\> \nonumber \\
&=& -\<v|S_a^\dagger \mathcal{O}_\alpha |v\> \cdot \<v| \mathcal{O}_\beta | v\> \nonumber \\
&=& \<v' | \mathcal{O}_\alpha v\> \cdot \<v| \mathcal{O}_\beta | v\> 
\end{eqnarray}
where the fourth equality follows from the fact that $v$ has short range correlations. On the other hand,
we know that $\<v'| \mathcal{O}_\alpha |v\> = 0$ by the special case discussed above. The relation
(\ref{lockd1}) follows immediately. 

Similarly, to derive (\ref{lockd2}), we note that
\begin{eqnarray}
\<v'|\mathcal{O} v'\> &=& \<v'|\mathcal{O}_\alpha \mathcal{O}_\beta v'\> \nonumber \\
&=& \<S_a v|\mathcal{O}_\alpha \mathcal{O}_\beta S_a v\> \nonumber \\
&=& \<v|S_a^\dagger \mathcal{O}_\alpha \mathcal{O}_\beta S_a v\> \nonumber \\
&=& \<v|(S_a^\dagger \mathcal{O}_\alpha S_a)(\mathcal{O}_\beta) |v\> \nonumber \\
&=& \<v| S_a^\dagger \mathcal{O}_\alpha S_a |v\> \cdot \<v|\mathcal{O}_\beta |v\> \nonumber \\
&=& \<v'|\mathcal{O}_\alpha |v'\> \cdot \<v|\mathcal{O}_\beta |v\> \\
\end{eqnarray}
Then, using the fact that $\<v'|\mathcal{O}_\alpha |v'\> = \<v| \mathcal{O}_\alpha |v\>$ we deduce
\begin{eqnarray}
\<v'|\mathcal{O} v'\> &=& \<v|\mathcal{O}_\alpha |v\> \cdot \<v|\mathcal{O}_\beta |v\> \nonumber \\
&=& \<v|\mathcal{O}_\alpha \mathcal{O}_\beta |v\> \nonumber \\
&=& \<v|\mathcal{O} v\>
\end{eqnarray}
This completes our proof of the local analogue of Kramers theorem.

\section{Adiabatic flux insertion and the edge theory} \label{fluxapp}
In this section, we analyze the effect of the adiabatic flux insertion of
$-1/e^*$ flux quanta using the edge theory (\ref{kedge}) to describe the
two edges of the cylinder. We show that the effect of this flux insertion process
is given by (\ref{fluxeffect}).

For now, we focus on one of the edges --
say the left edge. The adiabatic insertion of $-1/e^*$ flux quanta can be implemented
by applying a slowly varying vector potential $(A_t,A_x) = (0,f(t)/L)$
where $f(-\infty) = 0, f(\infty) = -2\pi/e^*$. Since the edge theory is quadratic,
we can analyze the effect of this flux insertion using the classical equations
of motion. We have
\begin{equation}
-\mathcal{K} \partial_t \partial_x \Phi + \mathcal{V} \partial_x^2 \Phi =
\tau (\partial_t A_x - \partial_x A_t)
\end{equation}
Integrating over $x$ and using the above expression for $A$, we derive
\begin{equation}
\frac{d\rho}{dt} = -\mathcal{K}^{-1} \frac{\tau}{2\pi} \frac{df}{dt}
\end{equation}
where $\rho_I \equiv \frac{1}{2\pi} \int dx \partial_x \Phi_I$.
The net effect of the flux insertion is therefore
\begin{equation}
\rho \rightarrow \rho + \frac{1}{e^*}\cdot \mathcal{K}^{-1} \tau
\label{Qeffect}
\end{equation}

On the other hand, if one quantizes (\ref{kedge}), one derives the canonical
commutation relations
\begin{equation}
[\Phi_I(y),\partial_x \Phi_J(x)] = 2\pi i \mathcal{K}^{-1}_{IJ} \delta(x-y)
\end{equation}
Letting $\Gamma(\Lambda) = \int \frac{dx}{\sqrt{L}} e^{i\Theta(\Lambda)}$
where $\Theta(\Lambda) = \Lambda^T \mathcal{K} \Phi$, we derive the commutation relation
\begin{equation}
[\rho, \Gamma(\Lambda)] = \Lambda \cdot \Gamma(\Lambda)
\label{Qcomm}
\end{equation}
Comparing (\ref{Qeffect}) and (\ref{Qcomm}), we conclude that the adiabatic flux insertion
is implemented by the operator $\Gamma(\frac{1}{e^*}\mathcal{K}^{-1} \tau) =
\Gamma(\Lambda_c)$.

Including the effect on the right edge as well, we
have
\begin{equation}
|\Psi_2\> = \Gamma_l(\Lambda_c)\Gamma_r(-\Lambda_c) |\Psi_1\>
\end{equation}
where $\Gamma_l, \Gamma_r$ denote the operators acting on the left
and right edges respectively. This completes our derivation of (\ref{fluxeffect}).

\bibliography{frtopinlong}

\end{document}